\documentclass[prl,floatfix,twocolumn,superscriptaddress,showpacs,preprintnumbers,longbibliography,nofootinbib]{revtex4-1}

\usepackage{amssymb,amsfonts,amsmath}
\usepackage{CJK} % for Chinese characters

\usepackage{float} 

\usepackage{hyperref}
\hypersetup{colorlinks=true, citecolor=blue, hypertexnames=false}
\usepackage[all]{hypcap} % let hyperlinks correctly point to figures rather than their captions; must be loaded after the hyperref package!

\usepackage{graphicx}% Include figure files
\usepackage{dcolumn}% Align table columns on decimal point
\usepackage{bm}% bold math
\usepackage{psfrag}
\usepackage{epsfig}
\usepackage{color}
\usepackage{bbm}
\usepackage[FIGTOPCAP,raggedright,nooneline]{subfigure}
\usepackage{verbatim}
\usepackage{upgreek} % for \tau

\usepackage{tikz}

\newcommand{\ignore}[1]{}

\newcommand{\eq}{Eq.\,}
\newcommand{\eqs}{Eqs.\,}
\newcommand{\fig}{Fig.\,}
\newcommand{\figs}{Figs.\,}
\newcommand{\cf} {cf.~}

\newcommand{\ie} {i.e.~}
\newcommand{\eg} {e.g.~}
\newcommand{\rref} {Ref.\,}

\newcommand{\figurepanel}[2]{\hyperref[#1]{\ref*{#1}(#2)}}

\makeatletter
\newcommand\footnoteref[1]{\protected@xdef\@thefnmark{\ref{#1}}\@footnotemark}
\makeatother

\usepackage{bibunits}

\newcommand{\nocontentsline}[3]{}
\newcommand{\tocless}[2]{\bgroup\let\addcontentsline=\nocontentsline#1{#2}\egroup}

\usepackage{filecontents}

\begin{document}
	\begin{bibunit}[apsrev4-1]
		\begin{CJK*}{UTF8}{} % instructed by  http://journals.aps.org/pra/authors/author-names-information
		\CJKfamily{bsmi}
	
	\title{
%	Generating dressed bound states in the continuum via multi-photon scattering and delayed quantum feedback OR\\
%	Using nonlinearity and delayed quantum feedback to excite a bound state in the continuum OR\\
	Exciting a Bound State in the Continuum through Multiphoton Scattering Plus Delayed Quantum Feedback }
	
	\author{Giuseppe Calaj\'o }
	\thanks{Present address: ICFO-Institut de Ciencies Fotoniques, 08860 Castelldefels (Barcelona), Spain. Contact email giuseppe.calajo@icfo.eu.}
	\affiliation{Vienna Center for Quantum Science and Technology, Atominstitut, TU Wien, Stadionallee 2, 1020 Vienna, Austria}
	
	\author{Yao-Lung L. Fang (方耀龍)}
	\affiliation{Department of Physics, Duke University, P.O. Box 90305, Durham, North Carolina 27708-0305, USA}
	\affiliation{Computational Science Initiative, Brookhaven National Laboratory, Upton, NY 11973, USA}
	\thanks{Present address}
    \affiliation{National Synchrotron Light Source II, Brookhaven National Laboratory, Upton, NY 11973, USA}
    \thanks{Present address}
	
	\author{Harold U. Baranger}
	\affiliation{Department of Physics, Duke University, P.O. Box 90305, Durham, North Carolina 27708-0305, USA}
	
	\author{Francesco Ciccarello}
	\affiliation{Department of Physics, Duke University, P.O. Box 90305, Durham, North Carolina 27708-0305, USA}
\affiliation{Universit$\grave{a}$ degli Studi di Palermo, Dipartimento di Fisica e Chimica, via Archirafi 36, I-90123 Palermo, Italia}

	\affiliation{NEST, Istituto Nanoscienze-CNR, Piazza S. Silvestro 12, 56127 Pisa, Italia}

	\date{January 13, 2019}
	
	%%%%
	\begin{abstract}
Excitation of a bound state in the continuum (BIC) through scattering is problematic since it is by definition uncoupled. Here, we consider a type of dressed BIC and show that it can be excited 
in a nonlinear system 
through multi-photon scattering and delayed quantum feedback. The system is a semi-infinite waveguide with linear dispersion coupled to a qubit, in which a single-photon, dressed BIC is known to exist. We show that this BIC can be populated via multi-photon scattering in the non-Markovian regime, where the photon delay time (due to the qubit-mirror distance) is comparable with the qubit's decay. A similar process excites the BIC existing in an infinite waveguide coupled to two distant qubits, thus yielding stationary entanglement between the qubits. This shows, in particular, that single-photon trapping via multi-photon scattering can occur without band-edge effects or cavities, the essential resource being instead the {\it delayed} quantum feedback provided by a single mirror or the emitters themselves.
	\end{abstract}
	
	% Leo: APS no longer uses PACS numbers...
%	\pacs{
%		42.50.Pq, %Cavity QED
%		42.50 Nn, %Quantum optical phenomena in absorbing, amplifying,
%		%dispersive and conducting media. Cooperative phenomena in Q.O. system
%		%03.65.Ge  %Solutions of wave equation:bound states
%		42.50.Ct 	%Quantum description of interaction of light and matter; related experiments
%		%42.65.An  42.50.Lc, 
%		%42.65., %Wi Nonlinear waveguides
%	}
	\maketitle
   \end{CJK*} % for Chinese characters

	\emph{Introduction.}---Waveguide Quantum ElectroDynamics (QED) is a 
%lively 
growing area of quantum optics investigating the coherent interaction between quantum emitters and the one-dimensional (1D) field of a waveguide \cite{LiaoPhyScr16,RoyRMP17,GuPR17}. In such systems, a growing number of unique nonlinear and interference phenomena are being unveiled, the occurrence of which typically relies on the 1D nature of such setups.
Among these is the formation of a class of 
%the long-sought bound-in-continuum states (BICs) 
bound states in the continuum (BIC), 
%\HUB{[HUB: I cut "long-sought" as there are lots of BICS seen  experimentally (water, acoustics, photonic slabs,...).]}
%BICs, first predicted by von Neumann and Wigner \cite{BICWigner}, 
which are bound stationary states that arise {\it within} a continuum of unbound states 
%(unlike usual bound states that occur in band gaps) 
\cite{HsuNRM16}. Topical questions are how to form and prepare such states so as to enable potential applications such as quantum memory, which requires light trapping at the few-photon level, of interest for quantum information processing \cite{KimbleNat08, qmreview, SaglamyurekNatPho18}.  
%Several efforts along this line have been made in various scenarios \cite{}. 
We show that addressing these questions involves studying 
%non-Markovian 
delayed quantum dynamics in the presence of nonlinearity. 
	
	An interesting class of BICs occurs in waveguide QED in the form of dressed states featuring one or more emitters, usually qubits, dressed with a {\it single} photon that is strictly confined within a finite region \cite{OrdonezPRA2006,Longhi,TanakaPRB07,TufarelliPRA13,GonzalezBallestroNJP13,RedchenkoPRA14,FacchiPRA16,FacchiJPC18}. The existence of such BICs relies on the {\it quantum feedback} provided by a mirror or the qubits themselves (since a qubit 
	%can effectively 
	behaves as a perfect mirror under 1D single-photon resonant scattering
	% Leo: we don't need that many refs here
	%\cite{ShenPRL05,ShenPRA07,ShenPRA09I,ChangNatPhy07}). 
	\cite{ShenPRL05,ChangNatPhy07}). 
	A natural way to populate these states is to excite the emitters and then let them decay: 
	% in vacuum:
	the system evolves towards the BIC with amplitude equal to the overlap between the BIC and the initial state. This results in incomplete decay of the emitter(s) and, in the case of two or more qubits, stationary entanglement \cite{FacchiPRA16,FacchiJPC18,TudelaPRL11,GonzalezTudelaPRL13,ChangNJP12}. As a hallmark, this approach for exciting 
	%dressed 
	BICs is most effective in the Markovian regime where the characteristic photonic {\it time delays}, denoted $\tau$, are very short (\eg the photon round-trip time between a qubit and mirror or between two qubits). 
	Indeed, as the time delay grows, the qubit component of the BIC decreases in favor of the photonic component
%	This is because the BIC's component having the qubit excited progressively shrinks for growing time delays in favor of a larger photonic component 
	\cite{TufarelliPRA13,GonzalezBallestroNJP13,RedchenkoPRA14}, making such decay-based schemes ineffective for 
	%long enough 
	large mirror-emitter or interemitter distances. This is a major limitation when entanglement creation is the 
	%main 
	goal \cite{GonzalezBallestroNJP13}.
	
	\begin{figure}
		\centering
		\includegraphics[width=75mm]{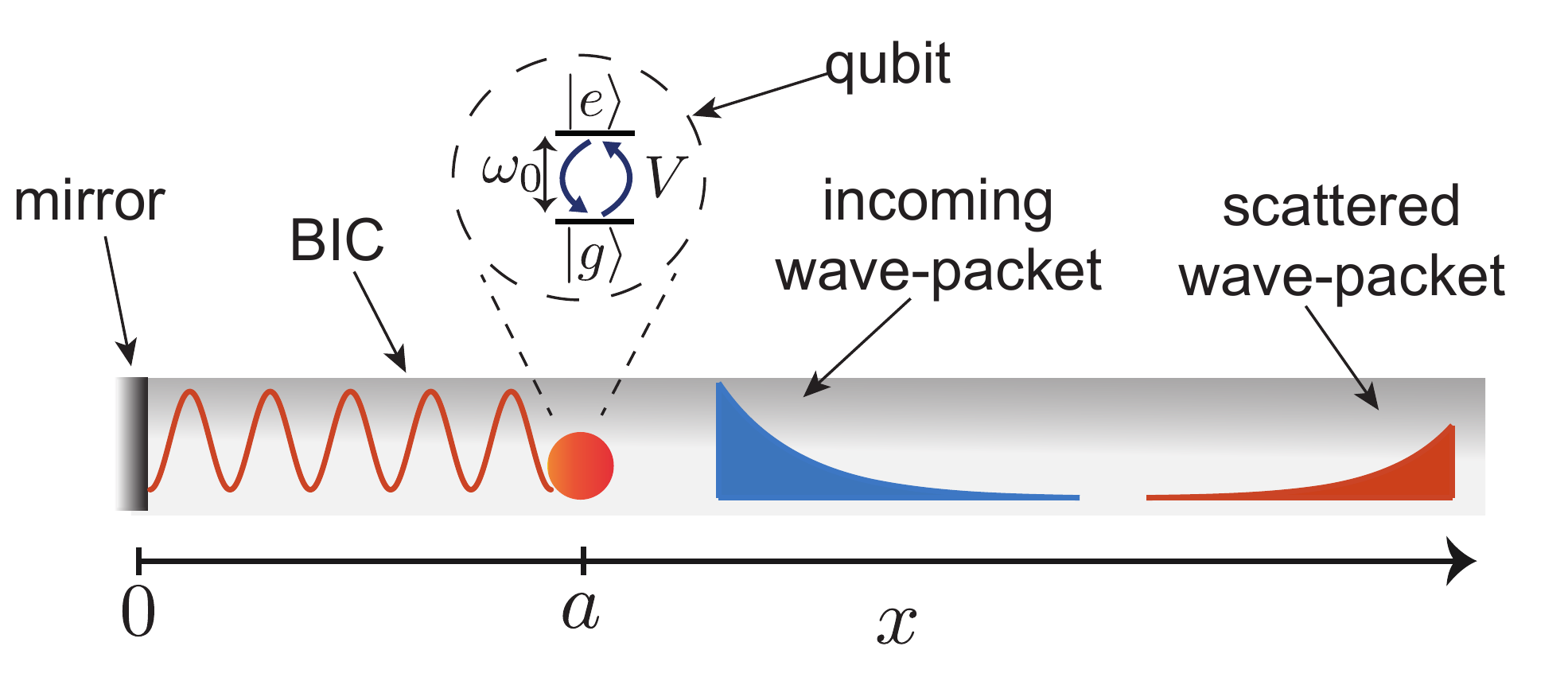}
		\caption{One-qubit setup: a {\it semi-infinite} waveguide, whose end lies at $x=0$ and 
acts as a perfect mirror,
%		embodies an effective perfect mirror, 
is coupled to a qubit at $x=a$. When a resonant standing wave can fit 
between the qubit and the mirror ($k_0a=m\pi$),
%in the qubit-mirror interspace (so $k_0a=m\pi$), there exists a possibility that 
an incoming two-photon wavepacket is {\it not} necessarily fully scattered off the qubit: a fraction remains trapped in the form of a dressed single-photon BIC. \label{Fig1_setup}}
\vspace*{-3mm}
	\end{figure}

	In order to generate such dressed BICs in the non-Markovian regime of significant time delays, one needs 
	%It appears that generating such dressed BICs in the non-Markovian regime of significant time delays is best accomplished by preparing 
	initial states that overlap the BIC's photonic component, which in practice calls for 
	%techniques based on 
	{\it photon scattering}. 
	%Based on this and the fact that the dressed BICs of concern here are of a single-photon nature, one might be tempted to expect that BICs excitation could be attained through single-photon scattering. 
	%Since BICs feature a field component one may wonder if they can be excited via photon scattering (with the QEs initially unexcited). 
	%Yet, a {\it single} photon scattered off the emitters cannot excite BICs since only unbound, single-excitation, scattering states are involved in such a process: the entire dynamics occurs within a sector of the Hilbert space orthogonal to the BIC. 
	A {\it single} photon scattered off the emitters cannot excite a BIC since the entire dynamics occurs in a sector of the Hilbert space orthogonal to the BIC.
	For {\it multi}-photon scattering, however, this argument does not hold because of the intrinsic qubit nonlinearity. 
	%and one can take advantage from the richer structure and energy degeneracy of the involved Hibert space sectors compared to the single-excitation subspace \cite{}. 
	Indeed, the role of two-photon scattering has been recognized previously \cite{LongoPRL10,LongoPRA11} in the context of exciting normal bound states (i.e., outside the continuum) that occur in cavity arrays coupled to qubits \cite{LombardoPRA14,CalajoPRA16,ShiPRX16,KocabasPRA16}.   
%%	Indeed, Longo {\it et al.}~\eg showed \cite{Longo,LongoPRA11} that {\it two-photon} scattering can excite the dressed bound states occurring {\it out} of the photonic continuum for an array of resonators coupled to a qubit \cite{Lombardo}, 
%	an effect relying on the structure of the energy spectrum 
%	%of dressed states 
%	in the two-excitation sector 
%	%of the Hilbert space 
%	\cite{CalajoPRA16}. 
	
	We show that dressed BICs in waveguide-QED setups can be excited via multi-photon scattering in two paradigmatic setups: a qubit coupled to a semi-infinite waveguide (see~\fig\ref{Fig1_setup}) and a pair of distant qubits coupled to an infinite waveguide [see~\fig\figurepanel{Fig4_2atom}{a}]. A perfectly sinusoidal photon wavefunction and stationary excitation of the emitters represents a clear signature of single-photon trapping. This provides a solvable example of non-Markovian quantum dynamics in a nonlinear system, a scenario of interest in many areas of contemporary physics \cite{ShuklaQPlasmasRMP11,GarciaMataJPA14,SwingleOtocNPhys18,RitschRMP13,AspelmeyerOptomechRMP14,WangNatComm17, RossiPRL18}.  
	
	{\it Model and BIC.}---Consider first a qubit coupled to the 1D field of a semi-infinite waveguide [\fig\figurepanel{Fig1_setup}{a}] having a linear dispersion  $\omega=v |k|$ (with $v$ the photon group velocity and $k$ the wavevector). The qubit's ground and excited states $|g\rangle$ and $|e\rangle$, respectively, are separated in energy by $\omega_0=vk_0$ (we set $\hbar=1$ throughout). 
	%, hence $k_0$ is the wavevector modulus of a photon resonant with the qubit. 
The end of the waveguide at $x=0$ is effectively a perfect mirror, while the qubit is placed at a distance $a$ from the mirror. The Hamiltonian under the rotating wave approximation (RWA) reads \cite{ShenPRL05,ShenPRA07,ShenPRA09I,ZhengPRA10,ZhengPRL13}
	\begin{align}\label{Hmodel}
	\hat H&=\omega_0\,\hat \sigma^\dagger\hat \sigma \!-iv\!\!\int_{0}^{\infty}\!\! \!{\rm d} x\!\left[\hat a^{\dagger}_R(x)\frac{d}{d x}\hat a_R(x)\!-\!\hat a^{\dagger}_L(x)\frac{d}{d x}\hat a_L(x)\right] \nonumber\\
	&\quad+ V\! \int_{0}^{\infty} \!\!\!\!{\rm d} x \left[\left(\hat a^{\dagger}_{ L}(x){+}\hat a^{\dagger}_{ R}(x)\right)\hat \sigma\!+\!{\rm H.c.}\right]\!\delta(x{-}a)\,,\!\!\!
	\end{align}
	with $\hat \sigma\!=|g\rangle\langle e|$, 
	$\hat a_{R(L)}(x)$ the bosonic field operator annihilating a right-going (left-going) photon at position $x$, and $V$ the atom-photon coupling. Due to the RWA, the total number of excitations $\hat N=\hat \sigma^\dagger\hat\sigma\!+\!\sum_{\eta=R,L}\int \!{\rm d} x\,\hat a_\eta^\dagger(x)\hat a_\eta(x)$ is conserved.
	
	In the single-excitation subspace ($N=1$), the spectrum of \eqref{Hmodel} comprises an infinite continuum of unbound dressed states $\{|\phi_k\rangle\}$ with energy $\omega_k\!=\!v|k|$ \cite{ShenPRL05,ShenPRA07,ShenPRA09I,ZhengPRA10,ZhengPRL13,GonzalezBallestroNJP13}, 
%interpretable %(in light of the Lippmann-Schwinger equations) 
each a scattering eigenstate in which an incoming photon is completely reflected.
% with 100\% probability. 
Notably, a further stationary state $|\phi_{b}\rangle$ exists when the 
condition $k_0a=m\pi$ (with $m=0, 1, \cdots$) is met. This BIC has the same energy $\omega_b=\omega_0$ as the qubit and is given by \cite{TufarelliPRA13,note}
	\begin{equation}
	|\phi_{b}\rangle{=} \varepsilon_b\left[\hat \sigma^\dag{\pm}i\sqrt{\tfrac{\Gamma}{2v}}\!\!\int_{0}^{a} \!\!{\rm d}x\left(e^{i k_0x}\hat a_R^{\dagger}(x){-}e^{-i k_0x}\hat a_L^{\dagger}(x)\right)\right]\!|g\rangle |0\rangle \label{phib}
	\end{equation}
with  $\Gamma=2 V^2/v$  the qubit's decay rate (without mirror). The qubit's excited-state population (referred to simply as ``population'' henceforth) is given by
	\begin{align}
	|\varepsilon_b|^2=\frac{1}{1+{\frac{1}{2}\, \Gamma\tau}}\,,\label{Pat}
%	|\varepsilon_b|^2=\frac{1}{1+\Gamma\tau/2}\,,\label{Pat}
	\end{align}
where  $\tau=2a/v$ is the {\it delay time}.
	%(photon's round-trip time between the qubit and mirror). 
\eqs\eqref{phib} and \eqref{Pat} fully specify the BIC.
% (up to an irrelevant global phase factor). BIC \eqref{phib} along with the set $\{|\phi_k\rangle\}$ form a complete basis of the single-excitation subspace of the Hilbert space.
	The photonic wavefunction has shape [we set $|x\rangle =\hat a^\dag (x)|0\rangle$ with $\hat a^\dag (x)=\left(\hat a_R^\dag (x)+\hat a_L^\dag (x)\right)$]
	\begin{align}\label{sine}
    \langle x|\phi_b\rangle\propto\sin(k_0 x)\,\,\,\,{\rm for}\,\,\,0\le x\le a\,,
	\end{align}
	while it vanishes at $x\not\in[0,a]\,$: the BIC is formed strictly between the qubit and the mirror, where the field profile is a pure sinusoid. When the BIC exists (i.e., for $k_0a=m\pi$)
% with $m>0$) 
the qubit does not fully decay in vacuum \cite{TufarelliPRA13,TufarelliPRA14,HoiNatPhy15}: 
since the overlap of the initial state $|e,0\rangle$ with the BIC is $\varepsilon_b$, $|\varepsilon_b|^2$ is also the probability of generating the BIC via vacuum decay. This probability decreases monotonically with 
%the rescaled \Leo{(``dimensionless'' should be better?)} 
delay time [\eq\eqref{Pat}], showing that vacuum decay is most effective when $\Gamma\tau$ is small. 
%Indeed, the initial state $|e,0\rangle$ overlaps the BIC by the amount $\langle e,0|\phi_b\rangle=\epsilon_b$, hence $|\epsilon_b|^2$ is also the probability of generating the BIC via vacuum decay. Based on \eq\eqref{Pat} this probability monotonically decreases with the rescaled \Leo{(``dimensionless'' should be better?)} delay time $\Gamma \tau$, showing that vacuum decay is most effective when $\Gamma\tau$ is small.
	%The emergence BIC can be understood as follows. Due to the hard-wall boundary condition imposed by the mirror at $x=0$, only sine modes of the field exist in the waveguide. Accordingly, the coupling stremgth between the qubit and a mode of wavevector $k$ is $g_k\propto \sin(ka)$. Under weak coupling ($\Gamma$ very small) the qubit interacts significantly only with modes such that $k\simeq k_0$. Thus, for $k_0a=m\pi$, $g_k\simeq 0$ and the decay is fully inhibited: $|e\rangle$ is a dark state. In this regime, $|\phi_b\rangle\simeq |e,0\rangle$. 
	%The BIC can be excited with probability $|\epsilon_b|^2$ by preparing the qubit in $|e\rangle$ and letting it decay (the decay being generally fractional)~\cite{Longhi,GonzalezBallestroNJP13,TufarelliPRA14,Pascazio}. 
	
	\emph{BIC generation scheme}.---Bound state \eqref{phib} cannot be generated, however, via single-photon scattering, 
%	however, since such a process 
which involves only the {unbound} states $\{|\phi_k\rangle\}$ that are all orthogonal to $|\phi_{b}\rangle$: during a transient time the photon may be absorbed by the qubit, but it is eventually fully released. We thus send a {\it two-photon} wavepacket 
%with the qubit initially unexcited 
such that the initial joint state is $|\Psi(0)\rangle = A %\tfrac{A}{\sqrt{2}} 
\iint_0^{\infty}\!{\rm d}x {\rm d}y\,  
	\left[\varphi^L_1(x)\varphi^L_2(y)+1\leftrightarrow2\right] 
	\hat a_L^{\dagger}(x)\hat a_L^{\dagger}(y)|g\rangle|0\rangle$, 
where $A$ is for normalization, $\varphi^L_i(x)$ is the wavefunction of a single left-propagating photon, and the qubit is not excited. 
The ensuing dynamics  in the two-excitation sector ($N=2$) 
is given by 
% hence at time $t$ the joint state has the form
	\begin{equation}\label{psit}
	\begin{split}
	&|\Psi(t)\rangle=
	\!\left[\sum_{\eta=R,L} \int_0^{\infty}\!\!{\rm d}x\, \psi_\eta(x,t)\hat a_\eta^{\dagger}(x)\hat\sigma^\dag\right.\\
	& \left.+\sum_{\eta,\eta'=R,L} \tfrac{1}{\sqrt{2}}\iint_0^{\infty}\!\! {\rm d}x {\rm d}y\,
	\chi_{\eta\eta'}(x,y,t)\hat a_{\eta}^{\dagger}(x)\hat a_{\eta'}^{\dagger}(y)\right]|g\rangle|0\rangle,
	\end{split}
	\end{equation}
	where $\chi_{\eta\eta'}(x,y,t)$ is the wavefunction of the two-photon component while $\psi_\eta(x,t)$ is the amplitude that the qubit is excited and a  right-(left-) propagating photon is found at position $x$.
% (initially, $\psi(x,0)=0$ and $\chi(x,y,0)=A\left[\varphi_1(x)\varphi_2(y)+1\leftrightarrow 2\right]$). 
%Based on \eqref{psit}, 
We define 
	\begin{equation}
	\begin{split}
	P_{\rm e}(t)\equiv &\sum_{\eta=R,L}\!\int_{0}^{\infty}\!\!{\rm d}x \,|\psi_\eta(x,t)|^2,\\
	P_{\rm ph}(t)\equiv & \;\;2 \!\!\!\sum_{\eta,\eta'=R,L}\int_0^a \!\!{\rm d}x\!\int_a^\infty \!\!\!{\rm d}y\,|\chi_{\eta\eta'}(x,y,t)|^2 \label{Pp-def}\!\
	\end{split}
	\end{equation}
	as, respectively, the qubit population and the probability that one photon lies in region $[0,a]$ and one in $(a,\infty)$.
	
	We first consider for simplicity a two-photon exponential wavepacket (sketched in \fig\ref{Fig1_setup}):
$\varphi^L_{1, 2}(x)=e^{-\Delta k|x-a|-ik_0(x-a)}\theta(x-a)$
where $v \Delta k$ is the bandwidth,
%wavevector width (hence $\Delta \omega{=}v \Delta k $ is the bandwidth) \Leo{(this symbol $\Delta \omega$ is never used, do we really need to define it?)} 
the carrier wavevector $k_0$ is resonant with the qubit, and the wavefront reaches the qubit at $t=0$. 
	In \fig\ref{Fig2_capture}, we plot results for the dynamics described by~\eqref{Hmodel} obtained numerically (for details see \cite{SupMat}). 
	%(for details see \cite{SupMat}).  
	%in a case such that BIC \eqref{phib} exists. 
% $\Gamma=\omega_0/20$, $v \Delta k =\Gamma/2$ (bandwidth equal to half the qubit's decay rate) and $k_0a{=}10\pi$ (ensuring the existence of the BIC). 
	As the wavepacket impinges on the qubit, its population $P_{\rm e}$ [\fig\figurepanel{Fig2_capture}{a}] exhibits a rise followed by a drop (photon absorption then re-emission) eventually converging to a small --- yet {\it finite} --- steady value. This shows that part of the excitation 
absorbed from the wavepacket is never released back. 
		\begin{figure}[t]
		\centering
		\includegraphics[width=85mm]{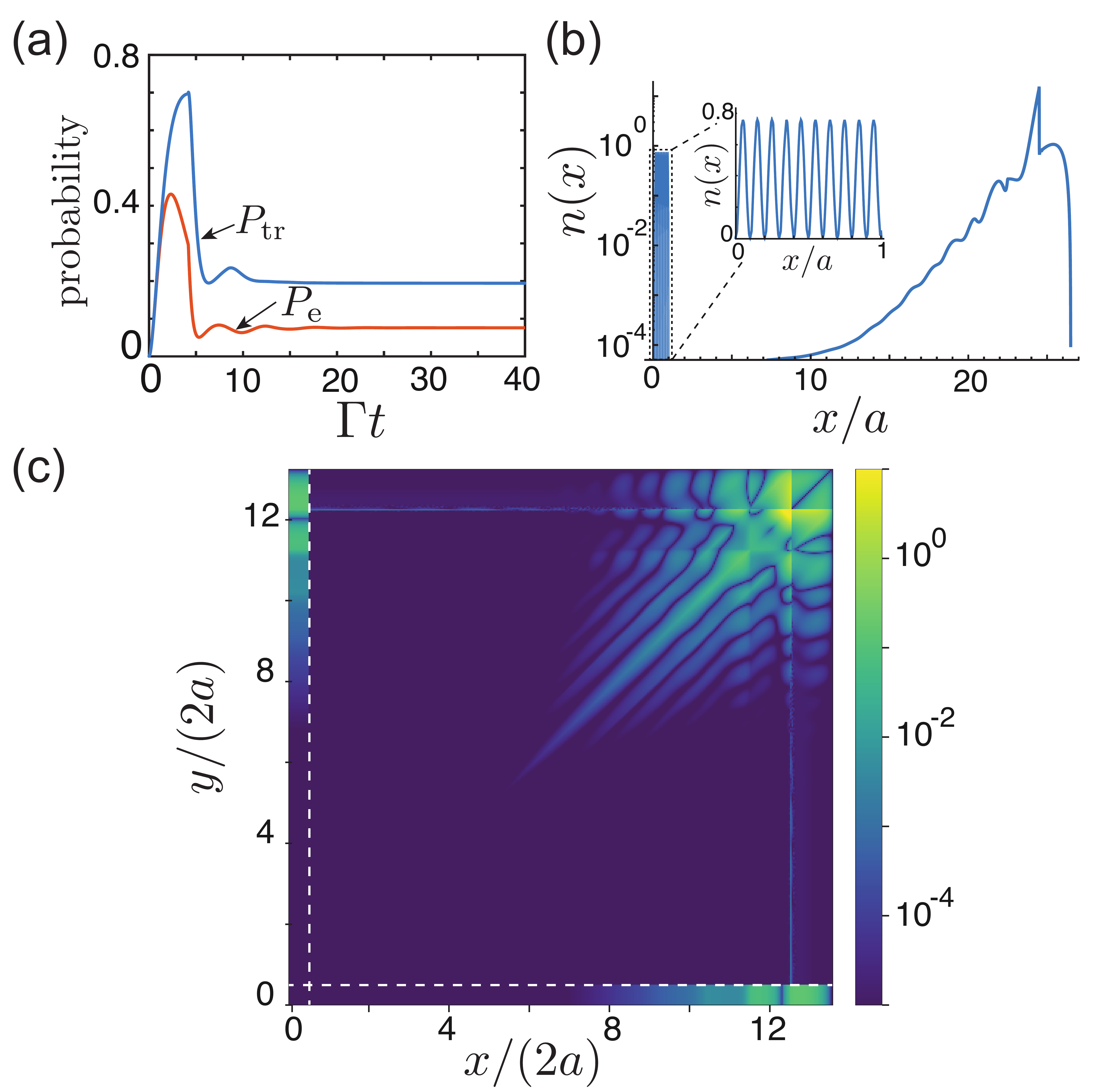}
		\caption{BIC generation via two-photon scattering. (a) Qubit population $P_{\rm e}$ and trapping probability $P_{\rm tr}$ as a function of time in units of $\Gamma^{-1}$. (b) Spatial profile of the field intensity $n(x)$ at the end of scattering. The inset highlights the sinusoidal wavefunction in the range $0\le x\le a$. (c) Two-photon probability density function $\sum_{\eta,\eta'=R,L}|\chi_{\eta\eta'}(x,y)|^2$  after scattering is complete ($t=t_f$). The white dashed lines $x{=}a$ and $y{=}a$ mark the qubit position.  
		[Panels (b) and (c) are plotted on a log scale and with $t_f{=}80/\Gamma$. We considered a two-photon exponential wavepacket with  $\Gamma\tau=\pi$, $k_0a=10\pi$ and $\Delta k =\Gamma/2v$.] 
			\label{Fig2_capture}
		}
	\end{figure}
	
The photon field in the same process is shown in 
\figs\figurepanel{Fig2_capture}{b} and \figurepanel{Fig2_capture}{c} displaying, respectively, 
%the spatial profile of the 
%[\cf\eq\eqref{psit}] 
 the field intensity $n(x)\!=\!\langle \Psi(t_f)|\hat a^\dag (x)\hat a (x)| \Psi(t_f)\rangle$ and the total two-photon probability density $\sum_{\eta,\eta'=R,L}|\chi_{\eta\eta'}(x,y,t_f)|^2$  at a time $t_f$ after the scattering process is complete. 
%($t_f\rightarrow\infty$). 
%long enough that the scattering process is over. 
The wavepacket is not entirely reflected back: a significant fraction remains trapped between the mirror and qubit, forming a perfectly sinusoidal wave with wavevector $k_0$ [\fig\figurepanel{Fig2_capture}{b}]. Remarkably, this stationary wave is of single-photon nature. Indeed, \fig\figurepanel{Fig2_capture}{c} shows that either both photons are reflected (top right corner) or one is scattered and the other remains trapped in the mirror-qubit interspace (top left and bottom right). Note that the probability that both photons are trapped (bottom left) is zero.
	\begin{figure}[h!]
		\centering
		\includegraphics[width=86mm]{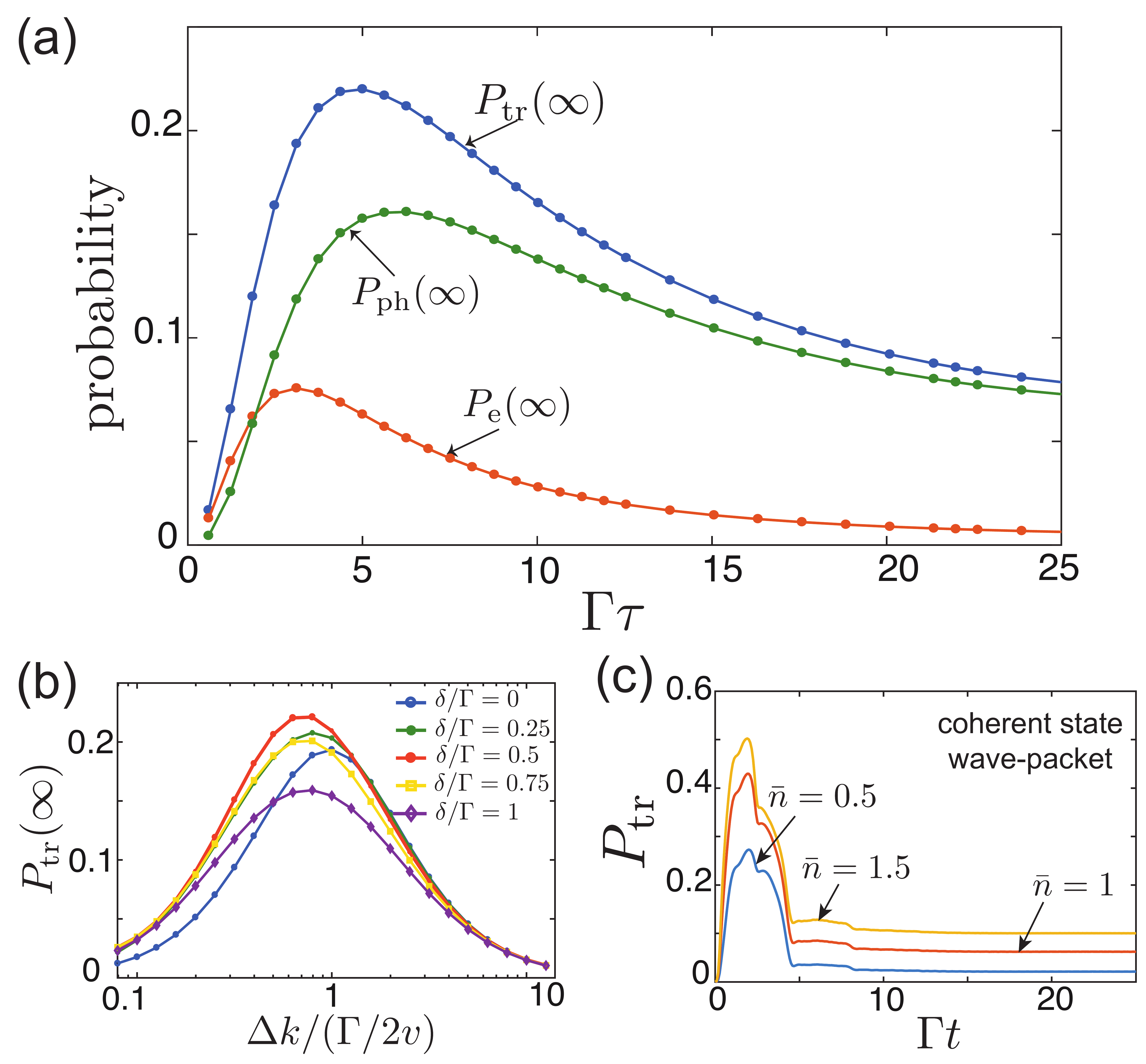}
		\caption{
			(a) Asymptotic values of $P_{\rm e}$, $P_{\rm ph}$, and $P_{\rm tr}$ as a function of the rescaled time delay $\Gamma \tau$ for $\delta=0$. At each point, $\Delta k $ is set so as to maximize $P_{\rm tr}$. 
			%at the considered value of $\Gamma\tau$. 
            Here we used the optimized $\Delta k $ shown in \cite{SupMat}.
			(b) Asymptotic value of $P_{\rm tr}(\infty)=P_{\rm BIC}$ against the wavepacket bandwidth for $\Gamma\tau=\pi$, $k_0a=10\pi$ and different values of detuning $\delta$, where we assumed that one photon has carrier wavevector $k_1=k_0{+}\delta/v$ and the other has $k_2=k_0{-}\delta/v$. 
			%In (b), we set $\delta=0$ and plot  
			(c) $P_{\rm tr}$ versus time for coherent-state wavepackets with the same shape. For computational reasons, only contributions up to three-photon Fock states are retained.  The parameters are the same as in \fig\ref{Fig2_capture}.
		    %(b)-(c) Dependence of our BIC generation scheme on (b) bandwidth and detuning and (c) time delay. In the incoming wavepacket, %The incoming wavepacket reads ${1}/{\sqrt{2}}\iint_0^{\infty}\!\! {\rm d}x {\rm d}y\,\,		\varphi_1(x)\varphi_2(y)\hat a^{\dagger}(x)\hat a^{\dagger}(y)|g\rangle|0\rangle$, where $\varphi_i(x)=e^{-\Delta k |x-a|-ik_i(x-a)}\theta(x{-}a)$ with $k_1=k_0{+}\delta/v$ and $k_2=k_0{-}\delta/v$. 
           \label{Fig3_optimization}}
	\end{figure}
	
	These outcomes, in light of the features of the BIC \eqref{phib}, suggest that, after scattering, the joint state has the form
	\begin{equation}\label{psif}
	\begin{split}
	|\Psi(t_f)\rangle&=\int_a^\infty \!\!{\rm d}x\,\xi_{R}(x,t_f)\hat a_R^\dag(x)|\phi_b\rangle\\
	&\,\,\,\,+\iint_a^{\infty}\!\! {\rm d}x {\rm d}y \,
	\beta_{RR}(x,y,t_f)\,\hat a_R^{\dagger}(x)\hat a_R^{\dagger}(y) |g,0\rangle\,,
	\end{split}
	\end{equation}
where in the first line a single photon has left the BIC region, 
%	with the first line describing a single photon leaving the interspace where the BIC \eqref{phib} arises, 
while the last line describes two outgoing photons.
%(Time $t$ in \eqref{psif} is such that the scattering is completely over.) 
%should be understood with $t\gg t_f$, i.e., the scattering is over.
	Let $P_{\rm tr}=P_{\rm e}+P_{\rm ph}$ 
	%[\cf\eq\eqref{Pp-def}] 
be the probability that either the qubit is excited or a photon is trapped between the mirror and qubit. It then follows from \eqref{psif} \cite{SupMat} that the asymptotic values of $P_{\rm tr}$ and $P_{\rm e}$ fulfill  
%	\begin{equation}
%	P_{\rm BIC}\!\equiv\!P_{\rm tr}(\infty)=\int_a^\infty \!\!\!{\rm d}x\,|\xi(x,\infty)|^2
%	=\left(1+{\tfrac{1}{2}\, \Gamma\tau} \right)P_{\rm e}(\infty),%\,,\!\!\!
%	%\!=\!\left(1+\tfrac{\Gamma}{2}\tau\right)P_{\rm e}(\infty),%\,,\!\!\!
%	\label{Ps}
%	\end{equation}
%where $P_{\rm BIC}$ is naturally interpreted as the probability of generating BIC \eqref{phib}. 
	\begin{equation}
	P_{\rm tr}(\infty)=\int_a^\infty \!\!\!{\rm d}x\,|\xi_R(x,\infty)|^2
	=\left(1+{\tfrac{1}{2}\, \Gamma\tau} \right)P_{\rm e}(\infty),
	\label{Ps}
	\end{equation}
which is naturally interpreted as the probability of generating the BIC,  $P_{\rm BIC}\equiv P_{\rm tr}(\infty)$. 
The time dependence of $P_{\rm tr}$ shown in \fig\figurepanel{Fig2_capture}{a} demonstrates that it reaches a finite steady value satisfying \eqref{Ps},
%that is larger than the $P_{\rm e}$'s one by the factor $(1{+}\Gamma\tau/2)$, in full agreement with the last identity in \eq\eqref{Ps}, 
confirming \eq\eqref{psif} and thus, the generation of the BIC. The identity \eqref{Ps} 
%$P_{\rm tr}(\infty){=}(1{+}\Gamma\tau/2)P_{\rm e}(\infty)$ 
was checked in all of the numerical results presented.
% in this work.

	{\it Dependence on time delay.}---A substantial delay time is essential for exciting the BIC. The parameter set in \fig\ref{Fig2_capture}, for instance, corresponds to $\Gamma \tau \simeq 3.14$. To highlight this dependence, we report in \fig\figurepanel{Fig3_optimization}{a} the steady state values of $P_{\rm e}$, $P_{\rm ph}$, and $P_{\rm tr}$, optimized with respect to $\Delta k $, as functions of $\Gamma \tau$.
	% in the two-photon scattering process. 
	%The parameters set in \fig\ref{Fig2_capture} correspond to $\Gamma \tau\simeq3.1$. A significant delay time is indeed essential for generating the BIC. This is shown in \fig\figurepanel{Fig3_optimization}{b} reporting the steady values of $P_{\rm e}$, $P_{\rm ph}$ and $P_{\rm tr}$ (optimized with respect to $\Delta k $) as functions of $\Gamma \tau$ in the two-photon scattering process. 
Both photon trapping and stationary qubit excitation are negligible in the Markovian regime $\Gamma\tau\ll1$, in sharp contrast to vacuum-decay schemes for which this is instead the optimal regime. A delay time $\Gamma \tau\gtrsim 1$ is required to make our BIC generation scheme effective; indeed, 
%In the considered range $0\le \Gamma\tau\le 25$, 
each of the three probabilities reaches a maximum at a delay of order $\Gamma\tau\sim 1$. Remarkably, $P_{\rm e}$ becomes negligible compared to $P_{\rm ph}$ for $\Gamma \tau\gtrsim 10$, showing that the photon component is dominant at large delays as expected from \eqs\eqref{phib} and \eqref{Pat}: In this regime, we thus get almost {\it pure} single-photon trapping
	
	%{The condition $\gamma\tau\sim1$, namely the need for a {\it delayed} quantum feedback due to the mirror, can be understood as follows. The two-level nature of the qubit imposes that it can absorb only one photon at a time. Moreover, in a one-dimensional waveguide the qubit behaves as a perfect mirror under single-photon scattering: after being absorbed and re-emitted by the qubit, an impinging photon is fully reflected back \cite{bibid}. Consider now a right-incoming two-photon wavepacket in the setup of \fig1 : only one of the two (dubbed 'photon A') is absorbed by the qubit, while the other one ('B') goes through propagating towards the mirror. While B reaches the mirror and is reflected back, the qubit re-emits photon A to the right. The key issue is now how long is the time taken by B to be back to the emitter's location, which is $\tau$ by definition, compared to the qubit's decay time $1/\gamma$.}
	
	{\it Dependence on bandwidth  and detuning.}---The efficiency of BIC generation depends on the width, $\Delta k $, of the injected wavepacket. In \fig\figurepanel{Fig3_optimization}{b} the optimal value is close to $\Gamma/2v$. 
%for the parameters of \fig\ref{Fig2_capture}. 
Thus, photon absorption is maximum when the wavepacket width is of order the qubit decay rate, in agreement with general expectations \cite{StobinskaEPL09,BaragiolaPRA12,FangNJP18}. The optimal $\Delta k $ as a function of delay time is given in the supplemental material \cite{SupMat}; for large $\Gamma\tau$, the optimal value saturates near $0.2\Gamma/v$. 
%	In \fig\ref{Fig2_capture}, we set $\Delta k {=}\Gamma/(2v)$. This value is indeed the optimal one \Leo{(only for $k_0a=10\pi$ with $\omega_0=20\Gamma$. For other cases the optimal choice will be different, see Supp Matt)} for generating the BIC when both photons are resonant with the qubit, as can be seen from \fig\figurepanel{Fig3_optimization}{a} showing $P_{\rm BIC}$ against $\Delta k $. This agrees with the general expectation (see \eg\rrefs\cite{StobinskaEPL09,BaragiolaPRA12,FangNJP18}) that photon absorption during the scattering transient is maximum when the wavepacket width is of the order of the qubit decay rate. 

Non-resonant photons can also be used to generate the BIC: results for a wavepacket of two photons detuned oppositely in energy are shown in 
%a blue-detuned photon and  red-detuned one 
\fig\figurepanel{Fig3_optimization}{b}. The optimal wavepacket width changes but remains of order $\Gamma$. As the detuning increases, the maximum $P_{\rm BIC}$ initially rises and then decays; 
%the optimal detuning for generating the BIC being $\delta=\Gamma/2$. 
note that the optimal detuning 
%for maximizing $P_{\rm BIC}$ 
is $\delta\approx\Gamma/2$.	At this value the nonlinear scattering flux was shown to peak~\cite{FangPE16,FangPRA17}, confirming that the intrinsic {\it nonlinearity} of the emitters is key to generating the BIC~\cite{SupMat}.

	{\it Coherent-state wavepacket.}---It is natural to wonder whether, instead of a two-photon pulse, the BIC can be excited using a  coherent-state wavepacket, which is easier to generate experimentally. 
%\marginpar{\HUB{answer to Leo in .tex}}
%\Leo{(I thought continuous driving is easier to generate? I could be wrong...)} 
%answer from Harold: Yes it is. But we need a finite density of photons, which implies a wavepacket because we can deal with only a few photons. It is an interesting question to ask what is the BIC amplitude under continuous driving such that there are a few photons in an interval of width inverse\Gamma. But I don't see how to address that question. 
In \fig\figurepanel{Fig3_optimization}{c} we consider the same setup, parameters, and wavepacket shape $\varphi(x)$ as in Fig.~\ref{Fig2_capture} but for a low-power coherent-state pulse \cite{ZhengPRA10} $|\alpha\rangle=e^{-|\alpha|^2}\sum_{n=0}^\infty(\alpha^n/n!)\left(\int{\rm d}x \varphi(x)\hat a_L^\dagger(x)\right)^n|g,0\rangle$ with the average photon number given by $\bar n=|\alpha|^2$. For $\bar n=1.5$, $P_{\rm tr}(\infty)$ is comparable to the one obtained with the two-photon pulse, demonstrating the effectiveness of using coherent states.

\begin{figure}[b]
	\centering
	\includegraphics[width=85mm]{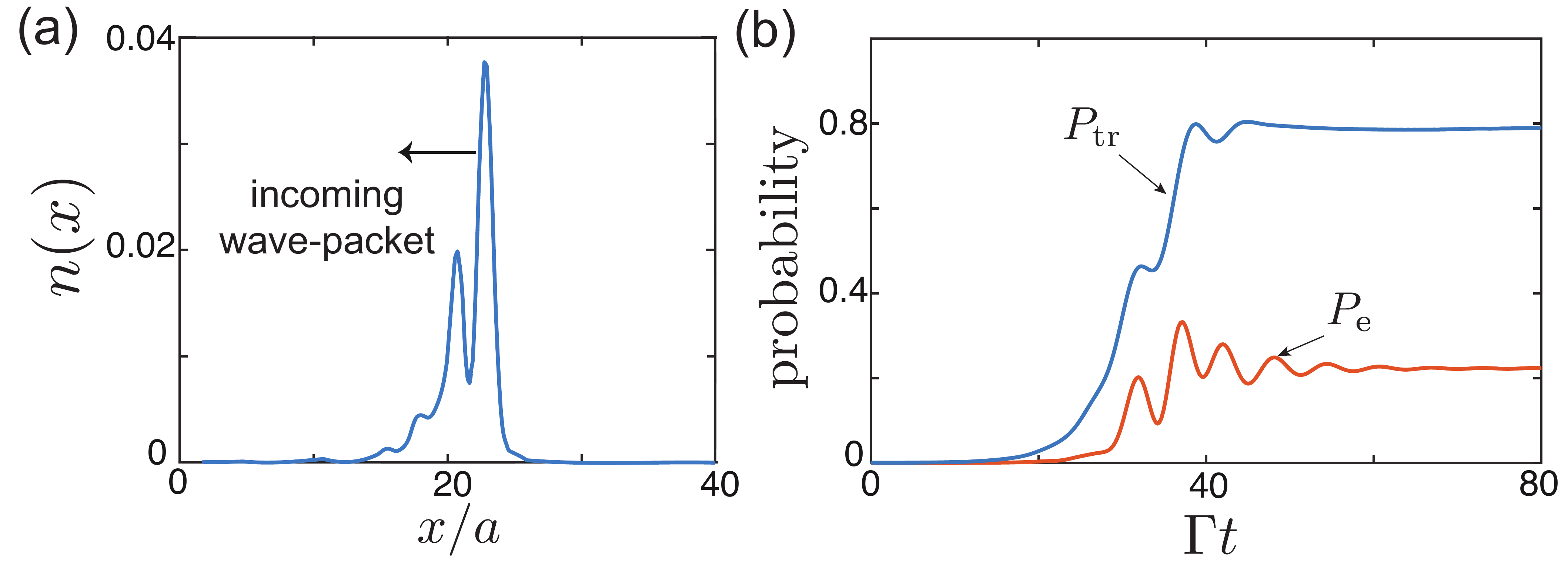}
	\caption{BIC generation scheme for the one-qubit setup using a structured-shape two-photon wavepacket (see \cite{SupMat}). (a) Photon density profile of the incoming wavepacket. (b) $P_{\rm BIC}$ and $P_{\rm e}$ versus time. 
		For this plot we fixed the distance to $k_0a=20\pi$ and the time delay to $\Gamma\tau=5$ to maximize the photon trapping probability [see \fig\figurepanel{Fig3_optimization}{b}].}
	\label{Fig5_BIC}
\end{figure}	
	
	{\it Increasing the BIC generation probability.}---We find that the trapping probability depends sensitively on the shape of the incoming wavepacket. While we have mostly used (\figs\ref{Fig2_capture}, \ref{Fig3_optimization},  \ref{Fig4_2atom}) the exponential pulse that is standard in the literature \cite{FangNJP18,RephaeliPRL12}, \fig\ref{Fig5_BIC} shows how engineering the wavepacket shape strongly enhances $P_{\rm BIC}$ \cite{CotrufoAluarXiv18}. 
We set here  $\Gamma\tau\!=\!5$, which roughly corresponds to the maximum of $P_{\rm tr}(\infty)=P_{\rm BIC}$ in \fig\figurepanel{Fig3_optimization}{a}. The engineered incoming two-photon wavepacket in \fig\figurepanel{Fig5_BIC}{a} (for methods
see \cite{SupMat}) yields $P_{\rm BIC}\simeq 80\%$, a value about four times larger. 
%	\Leo{(how is the wavepacket designed? will we explain in Supp Matt?)}
	%The corresponding probability to generate the BIC via vacuum decay [\cf\eq\eqref{Pat}] is $|\varepsilon_b|^2=0.28$, showing the effectiveness of the scattering-based BIC generation when time delays are significant.
	
	\begin{figure}[b]
		\centering
		\includegraphics[width=80mm]{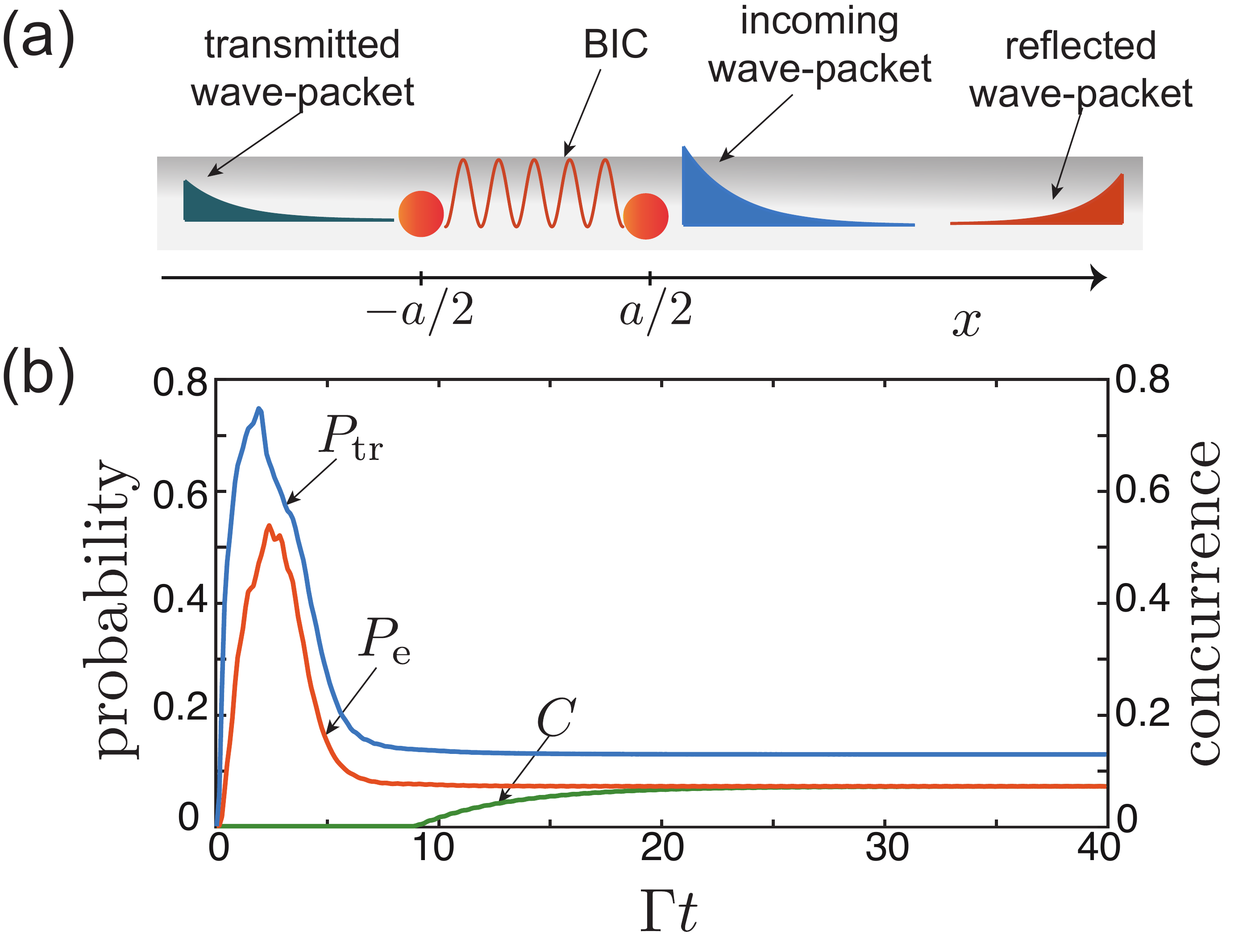}
		\caption{(a) Two-qubit setup: an {\it infinite }waveguide (no mirror) is coupled to a pair of qubits. (b) Probability to excite at least one qubit $P_{\rm e}$, trapping  probability $P_{\rm tr}$, and qubit-qubit concurrence $C$ versus time in a two-photon scattering process (see \cite{SupMat} for definition of $P_{\rm e}$, $P_{\rm tr}$ and $C$). 
			The wavepacket and parameters are the same as in \fig\ref{Fig2_capture}.
			The scheme generates a dressed BIC in a way analogous to the one-qubit setup in \fig\ref{Fig1_setup}, yielding however stationary entanglement between the qubits.
			\label{Fig4_2atom}
		}
	\end{figure}
	
	\emph{Two-qubit BIC.}--- A BIC very similar  to the one addressed above occurs in an infinite waveguide (no mirror) coupled to a pair of identical qubits~\cite{OrdonezPRA2006,TanakaPRB07,GonzalezBallestroNJP13,RedchenkoPRA14,FacchiPRA16,FacchiJPC18}. With the qubits placed at $x_{1}=- a/2$ and $x_{2}=a/2$ and for $k_0 a=m \pi$ [Fig.~\figurepanel{Fig4_2atom}{a}],
there exists a BIC given by 
%{\bf controllare $\Gamma$.}
	\begin{equation}\label{2atom_BIC}
		\begin{split}
	|\varphi_{b}\rangle = \varepsilon_b & \left[\hat \sigma_\pm^{\dagger}  - i\sqrt{\tfrac{\Gamma}{4v}}\!\int_{-a/2}^{a/2} \!\!{\rm d}x\left(e^{i k_0(x+a/2)}\hat a_R^{\dagger}(x)\right.\right. \\
	&\qquad\qquad  \left.\left. - e^{-i k_0(x+a/2)}\hat a_L^{\dagger}(x)\right)\right]\!|g_1,g_2\rangle|0\rangle,
	\end{split}
	\end{equation}
where now $|\varepsilon_b|^2 \!=\! 1/(1{+}\Gamma\tau/4)$,  
%and 
$\hat \sigma_\pm{=}(\hat \sigma_{1}{\pm}\hat \sigma_{2})/\sqrt{2}$,
and plus (minus) is used if $m$ is odd (even).  
%(with $\hat \sigma_{i}{=}|g\rangle_i\langle e|$) 
%\cite{note}. 
By tracing out the photonic field, \eq\eqref{2atom_BIC} clearly entails {\it entanglement} between the qubits. (In the familiar limit $\Gamma\tau\ll1$, for instance, the entangled state is  $\hat \sigma_\pm^\dag|g_1,g_2\rangle|0\rangle$, namely the sub- or super-radiant, maximally entangled state \cite{TudelaPRL11,ChangNJP12,ZhengPRL13,GonzalezTudelaPRL13,LalumierePRA13,vanLooScience13,FangPRA15}.)
	%of $a\ll 2\pi/k_0$
Thus, in the two-qubit setup of \fig\figurepanel{Fig4_2atom}{a}, our 	scattering-based approach to exciting the BIC can in particular generate entanglement. 
	
	This expectation is confirmed in \fig\figurepanel{Fig4_2atom}{b},   for which the same injected exponential wavepacket was used as in \fig\ref{Fig2_capture}. In addition to the probability to excite at least one qubit $P_{\rm e}$ and the probability to generate a BIC (\ref{2atom_BIC}), we plot the amount of entanglement between the qubits, as measured by the concurrence $C$ \cite{WoottersPRL98}. As for one qubit, the two-qubit BIC population reaches a steady value after scattering, resulting in an excitation stored in the qubits and hence stationary entanglement. Note the typical \cite{FicekPRA08} ``sudden birth'' of the entanglement.

	\emph{Conclusions.}---We have shown that dressed BICs occurring in waveguide-QED setups can be generated via multi-photon scattering. This enables single-photon capture and, for multiple emitters, production of stationary entanglement. These BICs differ significantly from purely excitonic subradiant states, as well as from BICs located entirely within the side-coupled quantum system, in that they involve the field of the waveguide itself. 
	
	For our method, it is critical to have \emph{nonlinear} emitters such as qubits; replacing them by bosonic modes, for example, will invalidate the whole scheme \cite{SupMat}.  
	%These two issues are addressed in \cite{SupMat}.
	
	While preparing this Letter we became aware of a related scheme by Cotrufo and Al\`{u} \cite{CotrufoAluarXiv18}. There however the BIC arises from a single system, comprised of a qubit and two cavities to provide feedback, side-coupled to an infinite waveguide. Here, instead, no cavities are present and the necessary quantum feedback is provided by a mirror [\cf\fig\ref{Fig1_setup}] or the emitters themselves [\cf\fig\figurepanel{Fig4_2atom}{a}]. Remarkably,  in order to generate the BIC, this feedback needs to be {\it delayed} [\cf\fig\figurepanel{Fig3_optimization}{a}]. 
	
	Investigating the non-Markovian effects of non-negligible delays is a new frontier of quantum optics \cite{TufarelliPRA13,RedchenkoPRA14,DornerPRA02,CarmelePRL13,ZhengPRL13,TufarelliPRA14,LaaksoPRL14,FangPRA15,GrimsmoPRL15,PichlerPRL16,RamosPRA16,TabakEPJQT16,WhalenQST17,GuimondQST17,PichlerPNAS17, RitschRMP13,AspelmeyerOptomechRMP14,WangNatComm17, GuoPRA17, RossiPRL18,ChalabiPRA18, TabakarXiv18,ChangarXiv18,TorrearXiv18,AnderssonarXiv18,FangNJP18}.
		 Here we have 
%highlighted a situation where one can 
taken advantage of such delays, demonstrating that their role can be constructive \cite{PichlerPNAS17,TorrearXiv18}. In particular, 
within the range considered, 
as shown in \fig\figurepanel{Fig3_optimization}{a}, 
long delays ($\Gamma\tau\gtrsim 20$) enable almost pure single-photon trapping (instead of hybrid atom-photon excitation). Remarkably, adding qubit losses, denoted by $\gamma_a$, makes the trapped photon decay slowly at the rate $\gamma_a|\varepsilon_b|^2$ \cite{SupMat}, suggesting that our scheme is more robust against emitter loss for larger delay $\tau$. 

Targets of ongoing investigation include exploring the regime of very long delays 
%($\Gamma \tau{\gg}1$) 
(beyond $\Gamma\tau\simeq 25$ in \fig\figurepanel{Fig3_optimization}{a} allowed by our current computational capabilities \cite{SupMat}) and deriving a systematic criterion to increase the generation probability by wavepacket engineering (possibly by exploiting time reversal symmetry \cite{StobinskaEPL09,CotrufoAluarXiv18}). We expect this line of research will become important to, for example, long-distance communication over quantum networks.
	
    %\tocless\acknowledgments
	We thank Peter Rabl, Shanhui Fan, and Tommaso Tufarelli for useful discussions. We acknowledge support from 
	the U.S.\,DOE Office of Science Division of Materials Sciences and Engineering under Grant No.\,DE-SC0005237 
	and the Fulbright Research Scholar Program. 
	The work at BNL was supported in part by BNL LDRD projects No.\,17-029 and No.\,19-002, 
	and New York State Urban Development Corporation, d/b/a Empire State Development, under contract No.\,AA289.
G.\,C.\ acknowledges support from  the Austrian Science Fund (FWF) through SFB FOQUS F40, DK CoQuS W 1210, and the START grant Y 591-N16. Part of the computation was completed using resources of the Center for Functional Nanomaterials, which is a U.S.\,DOE Office of Science Facility, and the Scientific Data and Computing Center, a component of the Computational Science Initiative, at Brookhaven National Laboratory under Contract No.\,DE-SC0012704.

%\bibliography{WQED,\jobname}

\begin{thebibliography}{70}%
\makeatletter
\providecommand \@ifxundefined [1]{%
 \@ifx{#1\undefined}
}%
\providecommand \@ifnum [1]{%
 \ifnum #1\expandafter \@firstoftwo
 \else \expandafter \@secondoftwo
 \fi
}%
\providecommand \@ifx [1]{%
 \ifx #1\expandafter \@firstoftwo
 \else \expandafter \@secondoftwo
 \fi
}%
\providecommand \natexlab [1]{#1}%
\providecommand \enquote  [1]{``#1''}%
\providecommand \bibnamefont  [1]{#1}%
\providecommand \bibfnamefont [1]{#1}%
\providecommand \citenamefont [1]{#1}%
\providecommand \href@noop [0]{\@secondoftwo}%
\providecommand \href [0]{\begingroup \@sanitize@url \@href}%
\providecommand \@href[1]{\@@startlink{#1}\@@href}%
\providecommand \@@href[1]{\endgroup#1\@@endlink}%
\providecommand \@sanitize@url [0]{\catcode `\\12\catcode `\$12\catcode
  `\&12\catcode `\#12\catcode `\^12\catcode `\_12\catcode `\%12\relax}%
\providecommand \@@startlink[1]{}%
\providecommand \@@endlink[0]{}%
\providecommand \url  [0]{\begingroup\@sanitize@url \@url }%
\providecommand \@url [1]{\endgroup\@href {#1}{\urlprefix }}%
\providecommand \urlprefix  [0]{URL }%
\providecommand \Eprint [0]{\href }%
\providecommand \doibase [0]{http://dx.doi.org/}%
\providecommand \selectlanguage [0]{\@gobble}%
\providecommand \bibinfo  [0]{\@secondoftwo}%
\providecommand \bibfield  [0]{\@secondoftwo}%
\providecommand \translation [1]{[#1]}%
\providecommand \BibitemOpen [0]{}%
\providecommand \bibitemStop [0]{}%
\providecommand \bibitemNoStop [0]{.\EOS\space}%
\providecommand \EOS [0]{\spacefactor3000\relax}%
\providecommand \BibitemShut  [1]{\csname bibitem#1\endcsname}%
\let\auto@bib@innerbib\@empty
%</preamble>
\bibitem [{\citenamefont {Liao}\ \emph {et~al.}(2016)\citenamefont {Liao},
  \citenamefont {Zeng}, \citenamefont {Nha},\ and\ \citenamefont
  {Zubairy}}]{LiaoPhyScr16}%
  \BibitemOpen
  \bibfield  {author} {\bibinfo {author} {\bibfnamefont {Z.}~\bibnamefont
  {Liao}}, \bibinfo {author} {\bibfnamefont {X.}~\bibnamefont {Zeng}}, \bibinfo
  {author} {\bibfnamefont {H.}~\bibnamefont {Nha}}, \ and\ \bibinfo {author}
  {\bibfnamefont {M.~S.}\ \bibnamefont {Zubairy}},\ }\bibfield  {title}
  {\enquote {\bibinfo {title} {{Photon transport in a one-dimensional
  nanophotonic waveguide QED system}},}\ }\href {\doibase
  10.1088/0031-8949/91/6/063004} {\bibfield  {journal} {\bibinfo  {journal}
  {Phys. Scr.}\ }\textbf {\bibinfo {volume} {91}},\ \bibinfo {pages} {063004}
  (\bibinfo {year} {2016})}\BibitemShut {NoStop}%
\bibitem [{\citenamefont {Roy}\ \emph {et~al.}(2017)\citenamefont {Roy},
  \citenamefont {Wilson},\ and\ \citenamefont {Firstenberg}}]{RoyRMP17}%
  \BibitemOpen
  \bibfield  {author} {\bibinfo {author} {\bibfnamefont {D.}~\bibnamefont
  {Roy}}, \bibinfo {author} {\bibfnamefont {C.~M.}\ \bibnamefont {Wilson}}, \
  and\ \bibinfo {author} {\bibfnamefont {O.}~\bibnamefont {Firstenberg}},\
  }\bibfield  {title} {\enquote {\bibinfo {title} {{Colloquium: Strongly
  interacting photons in one-dimensional continuum}},}\ }\href {\doibase
  10.1103/RevModPhys.89.021001} {\bibfield  {journal} {\bibinfo  {journal}
  {Rev. Mod. Phys.}\ }\textbf {\bibinfo {volume} {89}},\ \bibinfo {pages}
  {021001} (\bibinfo {year} {2017})}\BibitemShut {NoStop}%
\bibitem [{\citenamefont {Gu}\ \emph {et~al.}(2017)\citenamefont {Gu},
  \citenamefont {Kockum}, \citenamefont {Miranowicz}, \citenamefont {Liu},\
  and\ \citenamefont {Nori}}]{GuPR17}%
  \BibitemOpen
  \bibfield  {author} {\bibinfo {author} {\bibfnamefont {X.}~\bibnamefont
  {Gu}}, \bibinfo {author} {\bibfnamefont {A.~F.}\ \bibnamefont {Kockum}},
  \bibinfo {author} {\bibfnamefont {A.}~\bibnamefont {Miranowicz}}, \bibinfo
  {author} {\bibfnamefont {Y.~X.}\ \bibnamefont {Liu}}, \ and\ \bibinfo
  {author} {\bibfnamefont {F.}~\bibnamefont {Nori}},\ }\bibfield  {title}
  {\enquote {\bibinfo {title} {{Microwave photonics with superconducting
  quantum circuits}},}\ }\href {\doibase 10.1016/j.physrep.2017.10.002}
  {\bibfield  {journal} {\bibinfo  {journal} {Phys. Rep.}\ }\textbf {\bibinfo
  {volume} {718}},\ \bibinfo {pages} {1--102} (\bibinfo {year}
  {2017})}\BibitemShut {NoStop}%
\bibitem [{\citenamefont {Hsu}\ \emph {et~al.}(2016)\citenamefont {Hsu},
  \citenamefont {Zhen}, \citenamefont {Stone}, \citenamefont {Joannopoulos},\
  and\ \citenamefont {Solja{\v c}i{\'c}}}]{HsuNRM16}%
  \BibitemOpen
  \bibfield  {author} {\bibinfo {author} {\bibfnamefont {C.~W.}\ \bibnamefont
  {Hsu}}, \bibinfo {author} {\bibfnamefont {B.}~\bibnamefont {Zhen}}, \bibinfo
  {author} {\bibfnamefont {A.~D.}\ \bibnamefont {Stone}}, \bibinfo {author}
  {\bibfnamefont {J.~D.}\ \bibnamefont {Joannopoulos}}, \ and\ \bibinfo
  {author} {\bibfnamefont {M.}~\bibnamefont {Solja{\v c}i{\'c}}},\ }\bibfield
  {title} {\enquote {\bibinfo {title} {{Bound states in the continuum}},}\
  }\href {\doibase 10.1038/natrevmats.2016.48} {\bibfield  {journal} {\bibinfo
  {journal} {Nat. Rev. Mater.}\ }\textbf {\bibinfo {volume} {1}},\ \bibinfo
  {pages} {16048} (\bibinfo {year} {2016})}\BibitemShut {NoStop}%
\bibitem [{\citenamefont {Kimble}(2008)}]{KimbleNat08}%
  \BibitemOpen
  \bibfield  {author} {\bibinfo {author} {\bibfnamefont {H.~J.}\ \bibnamefont
  {Kimble}},\ }\bibfield  {title} {\enquote {\bibinfo {title} {The quantum
  internet},}\ }\href {\doibase 10.1038/nature07127} {\bibfield  {journal}
  {\bibinfo  {journal} {Nature}\ }\textbf {\bibinfo {volume} {453}},\ \bibinfo
  {pages} {1023} (\bibinfo {year} {2008})}\BibitemShut {NoStop}%
\bibitem [{\citenamefont {Lvovsky}\ \emph {et~al.}(2009)\citenamefont
  {Lvovsky}, \citenamefont {Sanders},\ and\ \citenamefont {Tittel}}]{qmreview}%
  \BibitemOpen
  \bibfield  {author} {\bibinfo {author} {\bibfnamefont {A.~I.}\ \bibnamefont
  {Lvovsky}}, \bibinfo {author} {\bibfnamefont {B.~C.}\ \bibnamefont
  {Sanders}}, \ and\ \bibinfo {author} {\bibfnamefont {W.}~\bibnamefont
  {Tittel}},\ }\bibfield  {title} {\enquote {\bibinfo {title} {Optical quantum
  memory},}\ }\href {\doibase 10.1038/nphoton.2009.231} {\bibfield  {journal}
  {\bibinfo  {journal} {Nat. Photonics}\ }\textbf {\bibinfo {volume} {3}},\
  \bibinfo {pages} {706} (\bibinfo {year} {2009})}\BibitemShut {NoStop}%
\bibitem [{\citenamefont {Saglamyurek}\ \emph {et~al.}(2018)\citenamefont
  {Saglamyurek}, \citenamefont {Hrushevskyi}, \citenamefont {Rastogi},
  \citenamefont {Heshami},\ and\ \citenamefont
  {LeBlanc}}]{SaglamyurekNatPho18}%
  \BibitemOpen
  \bibfield  {author} {\bibinfo {author} {\bibfnamefont {E.}~\bibnamefont
  {Saglamyurek}}, \bibinfo {author} {\bibfnamefont {T.}~\bibnamefont
  {Hrushevskyi}}, \bibinfo {author} {\bibfnamefont {A.}~\bibnamefont
  {Rastogi}}, \bibinfo {author} {\bibfnamefont {K.}~\bibnamefont {Heshami}}, \
  and\ \bibinfo {author} {\bibfnamefont {L.~J.}\ \bibnamefont {LeBlanc}},\
  }\bibfield  {title} {\enquote {\bibinfo {title} {{Coherent storage and
  manipulation of broadband photons via dynamically controlled
  Autler{\textendash}Townes splitting}},}\ }\href@noop {} {\bibfield  {journal}
  {\bibinfo  {journal} {Nat. Photonics}\ } (\bibinfo {year}
  {2018})}\BibitemShut {NoStop}%
\bibitem [{\citenamefont {Ordonez}\ \emph {et~al.}(2006)\citenamefont
  {Ordonez}, \citenamefont {Na},\ and\ \citenamefont {Kim}}]{OrdonezPRA2006}%
  \BibitemOpen
  \bibfield  {author} {\bibinfo {author} {\bibfnamefont {G.}~\bibnamefont
  {Ordonez}}, \bibinfo {author} {\bibfnamefont {K.}~\bibnamefont {Na}}, \ and\
  \bibinfo {author} {\bibfnamefont {S.}~\bibnamefont {Kim}},\ }\bibfield
  {title} {\enquote {\bibinfo {title} {Bound states in the continuum in
  quantum-dot pairs},}\ }\href {\doibase 10.1103/PhysRevA.73.022113} {\bibfield
   {journal} {\bibinfo  {journal} {Phys. Rev. A}\ }\textbf {\bibinfo {volume}
  {73}},\ \bibinfo {pages} {022113} (\bibinfo {year} {2006})}\BibitemShut
  {NoStop}%
\bibitem [{\citenamefont {Longhi}(2007)}]{Longhi}%
  \BibitemOpen
  \bibfield  {author} {\bibinfo {author} {\bibfnamefont {S.}~\bibnamefont
  {Longhi}},\ }\bibfield  {title} {\enquote {\bibinfo {title} {{Bound states in
  the continuum in a single-level Fano-Anderson model}},}\ }\href {\doibase
  10.1140/epjb/e2007-00143-2} {\bibfield  {journal} {\bibinfo  {journal} {Eur.
  Phys. J. B}\ }\textbf {\bibinfo {volume} {57}},\ \bibinfo {pages} {45--51}
  (\bibinfo {year} {2007})}\BibitemShut {NoStop}%
\bibitem [{\citenamefont {Tanaka}\ \emph {et~al.}(2007)\citenamefont {Tanaka},
  \citenamefont {Garmon}, \citenamefont {Ordonez},\ and\ \citenamefont
  {Petrosky}}]{TanakaPRB07}%
  \BibitemOpen
  \bibfield  {author} {\bibinfo {author} {\bibfnamefont {S.}~\bibnamefont
  {Tanaka}}, \bibinfo {author} {\bibfnamefont {S.}~\bibnamefont {Garmon}},
  \bibinfo {author} {\bibfnamefont {G.}~\bibnamefont {Ordonez}}, \ and\
  \bibinfo {author} {\bibfnamefont {T.}~\bibnamefont {Petrosky}},\ }\bibfield
  {title} {\enquote {\bibinfo {title} {Electron trapping in a one-dimensional
  semiconductor quantum wire with multiple impurities},}\ }\href {\doibase
  10.1103/PhysRevB.76.153308} {\bibfield  {journal} {\bibinfo  {journal} {Phys.
  Rev. B}\ }\textbf {\bibinfo {volume} {76}},\ \bibinfo {pages} {153308}
  (\bibinfo {year} {2007})}\BibitemShut {NoStop}%
\bibitem [{\citenamefont {Tufarelli}\ \emph {et~al.}(2013)\citenamefont
  {Tufarelli}, \citenamefont {Ciccarello},\ and\ \citenamefont
  {Kim}}]{TufarelliPRA13}%
  \BibitemOpen
  \bibfield  {author} {\bibinfo {author} {\bibfnamefont {T.}~\bibnamefont
  {Tufarelli}}, \bibinfo {author} {\bibfnamefont {F.}~\bibnamefont
  {Ciccarello}}, \ and\ \bibinfo {author} {\bibfnamefont {M.~S.}\ \bibnamefont
  {Kim}},\ }\bibfield  {title} {\enquote {\bibinfo {title} {Dynamics of
  spontaneous emission in a single-end photonic waveguide},}\ }\href {\doibase
  10.1103/PhysRevA.87.013820} {\bibfield  {journal} {\bibinfo  {journal} {Phys.
  Rev. A}\ }\textbf {\bibinfo {volume} {87}},\ \bibinfo {pages} {013820}
  (\bibinfo {year} {2013})}\BibitemShut {NoStop}%
\bibitem [{\citenamefont {Gonzalez-Ballestero}\ \emph
  {et~al.}(2013)\citenamefont {Gonzalez-Ballestero}, \citenamefont
  {Garcia-Vidal},\ and\ \citenamefont {Moreno}}]{GonzalezBallestroNJP13}%
  \BibitemOpen
  \bibfield  {author} {\bibinfo {author} {\bibfnamefont {C.}~\bibnamefont
  {Gonzalez-Ballestero}}, \bibinfo {author} {\bibfnamefont {F.~J.}\
  \bibnamefont {Garcia-Vidal}}, \ and\ \bibinfo {author} {\bibfnamefont
  {E.}~\bibnamefont {Moreno}},\ }\bibfield  {title} {\enquote {\bibinfo {title}
  {Non-{M}arkovian effects in waveguide-mediated entanglement},}\ }\href
  {\doibase 10.1088/1367-2630/15/7/073015} {\bibfield  {journal} {\bibinfo
  {journal} {New J. Phys.}\ }\textbf {\bibinfo {volume} {15}},\ \bibinfo
  {pages} {073015} (\bibinfo {year} {2013})}\BibitemShut {NoStop}%
\bibitem [{\citenamefont {Redchenko}\ and\ \citenamefont
  {Yudson}(2014)}]{RedchenkoPRA14}%
  \BibitemOpen
  \bibfield  {author} {\bibinfo {author} {\bibfnamefont {E.~S.}\ \bibnamefont
  {Redchenko}}\ and\ \bibinfo {author} {\bibfnamefont {V.~I.}\ \bibnamefont
  {Yudson}},\ }\bibfield  {title} {\enquote {\bibinfo {title} {Decay of
  metastable excited states of two qubits in a waveguide},}\ }\href {\doibase
  10.1103/PhysRevA.90.063829} {\bibfield  {journal} {\bibinfo  {journal} {Phys.
  Rev. A}\ }\textbf {\bibinfo {volume} {90}},\ \bibinfo {pages} {063829}
  (\bibinfo {year} {2014})}\BibitemShut {NoStop}%
\bibitem [{\citenamefont {Facchi}\ \emph {et~al.}(2016)\citenamefont {Facchi},
  \citenamefont {Kim}, \citenamefont {Pascazio}, \citenamefont {Pepe},
  \citenamefont {Pomarico},\ and\ \citenamefont {Tufarelli}}]{FacchiPRA16}%
  \BibitemOpen
  \bibfield  {author} {\bibinfo {author} {\bibfnamefont {P.}~\bibnamefont
  {Facchi}}, \bibinfo {author} {\bibfnamefont {M.~S.}\ \bibnamefont {Kim}},
  \bibinfo {author} {\bibfnamefont {S.}~\bibnamefont {Pascazio}}, \bibinfo
  {author} {\bibfnamefont {F.~V.}\ \bibnamefont {Pepe}}, \bibinfo {author}
  {\bibfnamefont {D.}~\bibnamefont {Pomarico}}, \ and\ \bibinfo {author}
  {\bibfnamefont {T.}~\bibnamefont {Tufarelli}},\ }\bibfield  {title} {\enquote
  {\bibinfo {title} {{Bound states and entanglement generation in waveguide
  quantum electrodynamics}},}\ }\href {\doibase 10.1103/PhysRevA.94.043839}
  {\bibfield  {journal} {\bibinfo  {journal} {Phys. Rev. A}\ }\textbf {\bibinfo
  {volume} {94}},\ \bibinfo {pages} {043839} (\bibinfo {year}
  {2016})}\BibitemShut {NoStop}%
\bibitem [{\citenamefont {Facchi}\ \emph {et~al.}(2018)\citenamefont {Facchi},
  \citenamefont {Pascazio}, \citenamefont {Pepe},\ and\ \citenamefont
  {Yuasa}}]{FacchiJPC18}%
  \BibitemOpen
  \bibfield  {author} {\bibinfo {author} {\bibfnamefont {P.}~\bibnamefont
  {Facchi}}, \bibinfo {author} {\bibfnamefont {S.}~\bibnamefont {Pascazio}},
  \bibinfo {author} {\bibfnamefont {F.~V.}\ \bibnamefont {Pepe}}, \ and\
  \bibinfo {author} {\bibfnamefont {K.}~\bibnamefont {Yuasa}},\ }\bibfield
  {title} {\enquote {\bibinfo {title} {Long-lived entanglement of two
  multilevel atoms in a waveguide},}\ }\href
  {http://stacks.iop.org/2399-6528/2/i=3/a=035006} {\bibfield  {journal}
  {\bibinfo  {journal} {J. Phys. Commun.}\ }\textbf {\bibinfo {volume} {2}},\
  \bibinfo {pages} {035006} (\bibinfo {year} {2018})}\BibitemShut {NoStop}%
\bibitem [{\citenamefont {Shen}\ and\ \citenamefont {Fan}(2005)}]{ShenPRL05}%
  \BibitemOpen
  \bibfield  {author} {\bibinfo {author} {\bibfnamefont {J.~T.}\ \bibnamefont
  {Shen}}\ and\ \bibinfo {author} {\bibfnamefont {S.}~\bibnamefont {Fan}},\
  }\bibfield  {title} {\enquote {\bibinfo {title} {Coherent single photon
  transport in a one-dimensional waveguide coupled with superconducting quantum
  bits},}\ }\href {\doibase 10.1103/PhysRevLett.95.213001} {\bibfield
  {journal} {\bibinfo  {journal} {Phys. Rev. Lett.}\ }\textbf {\bibinfo
  {volume} {95}},\ \bibinfo {pages} {213001} (\bibinfo {year}
  {2005})}\BibitemShut {NoStop}%
\bibitem [{\citenamefont {Chang}\ \emph {et~al.}(2007)\citenamefont {Chang},
  \citenamefont {S\o{}rensen}, \citenamefont {Demler},\ and\ \citenamefont
  {Lukin}}]{ChangNatPhy07}%
  \BibitemOpen
  \bibfield  {author} {\bibinfo {author} {\bibfnamefont {D.~E.}\ \bibnamefont
  {Chang}}, \bibinfo {author} {\bibfnamefont {A.~S.}\ \bibnamefont
  {S\o{}rensen}}, \bibinfo {author} {\bibfnamefont {E.~A.}\ \bibnamefont
  {Demler}}, \ and\ \bibinfo {author} {\bibfnamefont {M.~D.}\ \bibnamefont
  {Lukin}},\ }\bibfield  {title} {\enquote {\bibinfo {title} {A single-photon
  transistor using nanoscale surface plasmons},}\ }\href {\doibase
  10.1038/nphys708} {\bibfield  {journal} {\bibinfo  {journal} {Nature Phys.}\
  }\textbf {\bibinfo {volume} {3}},\ \bibinfo {pages} {807--812} (\bibinfo
  {year} {2007})}\BibitemShut {NoStop}%
\bibitem [{\citenamefont {Gonz{\'a}lez-Tudela}\ \emph
  {et~al.}(2011)\citenamefont {Gonz{\'a}lez-Tudela}, \citenamefont
  {Martin-Cano}, \citenamefont {Moreno}, \citenamefont {Mart{\'\i}n-Moreno},\
  and\ \citenamefont {Tejedor}}]{TudelaPRL11}%
  \BibitemOpen
  \bibfield  {author} {\bibinfo {author} {\bibfnamefont {A.}~\bibnamefont
  {Gonz{\'a}lez-Tudela}}, \bibinfo {author} {\bibfnamefont {D.}~\bibnamefont
  {Martin-Cano}}, \bibinfo {author} {\bibfnamefont {E.}~\bibnamefont {Moreno}},
  \bibinfo {author} {\bibfnamefont {L.}~\bibnamefont {Mart{\'\i}n-Moreno}}, \
  and\ \bibinfo {author} {\bibfnamefont {F.}~\bibnamefont {Tejedor},
  \bibfnamefont {C./Garc{\'\i}a-Vidal}},\ }\bibfield  {title} {\enquote
  {\bibinfo {title} {Entanglement of two qubits mediated by one-dimensional
  plasmonic waveguides},}\ }\href {\doibase 10.1103/PhysRevLett.106.020501}
  {\bibfield  {journal} {\bibinfo  {journal} {Phys. Rev. Lett.}\ }\textbf
  {\bibinfo {volume} {106}},\ \bibinfo {pages} {020501} (\bibinfo {year}
  {2011})}\BibitemShut {NoStop}%
\bibitem [{\citenamefont {Gonz{\'a}lez-Tudela}\ and\ \citenamefont
  {Porras}(2013)}]{GonzalezTudelaPRL13}%
  \BibitemOpen
  \bibfield  {author} {\bibinfo {author} {\bibfnamefont {A.}~\bibnamefont
  {Gonz{\'a}lez-Tudela}}\ and\ \bibinfo {author} {\bibfnamefont
  {D.}~\bibnamefont {Porras}},\ }\bibfield  {title} {\enquote {\bibinfo {title}
  {Mesoscopic entanglement induced by spontaneous emission in solid-state
  quantum optics},}\ }\href {\doibase 10.1103/PhysRevLett.110.080502}
  {\bibfield  {journal} {\bibinfo  {journal} {Phys. Rev. Lett.}\ }\textbf
  {\bibinfo {volume} {110}},\ \bibinfo {pages} {080502} (\bibinfo {year}
  {2013})}\BibitemShut {NoStop}%
\bibitem [{\citenamefont {Chang}\ \emph {et~al.}(2012)\citenamefont {Chang},
  \citenamefont {Jiang}, \citenamefont {Gorshkov},\ and\ \citenamefont
  {Kimble}}]{ChangNJP12}%
  \BibitemOpen
  \bibfield  {author} {\bibinfo {author} {\bibfnamefont {D.~E.}\ \bibnamefont
  {Chang}}, \bibinfo {author} {\bibfnamefont {L.}~\bibnamefont {Jiang}},
  \bibinfo {author} {\bibfnamefont {A.~V.}\ \bibnamefont {Gorshkov}}, \ and\
  \bibinfo {author} {\bibfnamefont {H.~J.}\ \bibnamefont {Kimble}},\ }\bibfield
   {title} {\enquote {\bibinfo {title} {Cavity {QED} with atomic mirrors},}\
  }\href {\doibase 10.1088/1367-2630/14/6/063003} {\bibfield  {journal}
  {\bibinfo  {journal} {New J. Phys.}\ }\textbf {\bibinfo {volume} {14}},\
  \bibinfo {pages} {063003} (\bibinfo {year} {2012})}\BibitemShut {NoStop}%
\bibitem [{\citenamefont {Longo}\ \emph {et~al.}(2010)\citenamefont {Longo},
  \citenamefont {Schmitteckert},\ and\ \citenamefont {Busch}}]{LongoPRL10}%
  \BibitemOpen
  \bibfield  {author} {\bibinfo {author} {\bibfnamefont {P.}~\bibnamefont
  {Longo}}, \bibinfo {author} {\bibfnamefont {P.}~\bibnamefont
  {Schmitteckert}}, \ and\ \bibinfo {author} {\bibfnamefont {K.}~\bibnamefont
  {Busch}},\ }\bibfield  {title} {\enquote {\bibinfo {title} {Few-photon
  transport in low-dimensional systems: Interaction-induced radiation
  trapping},}\ }\href {\doibase 10.1103/PhysRevLett.104.023602} {\bibfield
  {journal} {\bibinfo  {journal} {Phys. Rev. Lett.}\ }\textbf {\bibinfo
  {volume} {104}},\ \bibinfo {pages} {023602} (\bibinfo {year}
  {2010})}\BibitemShut {NoStop}%
\bibitem [{\citenamefont {Longo}\ \emph {et~al.}(2011)\citenamefont {Longo},
  \citenamefont {Schmitteckert},\ and\ \citenamefont {Busch}}]{LongoPRA11}%
  \BibitemOpen
  \bibfield  {author} {\bibinfo {author} {\bibfnamefont {P.}~\bibnamefont
  {Longo}}, \bibinfo {author} {\bibfnamefont {P.}~\bibnamefont
  {Schmitteckert}}, \ and\ \bibinfo {author} {\bibfnamefont {K.}~\bibnamefont
  {Busch}},\ }\bibfield  {title} {\enquote {\bibinfo {title} {Few-photon
  transport in low-dimensional systems},}\ }\href {\doibase
  10.1103/PhysRevA.83.063828} {\bibfield  {journal} {\bibinfo  {journal} {Phys.
  Rev. A}\ }\textbf {\bibinfo {volume} {83}},\ \bibinfo {pages} {063828}
  (\bibinfo {year} {2011})}\BibitemShut {NoStop}%
\bibitem [{\citenamefont {Lombardo}\ \emph {et~al.}(2014)\citenamefont
  {Lombardo}, \citenamefont {Ciccarello},\ and\ \citenamefont
  {Palma}}]{LombardoPRA14}%
  \BibitemOpen
  \bibfield  {author} {\bibinfo {author} {\bibfnamefont {F.}~\bibnamefont
  {Lombardo}}, \bibinfo {author} {\bibfnamefont {F.}~\bibnamefont
  {Ciccarello}}, \ and\ \bibinfo {author} {\bibfnamefont {G.~M.}\ \bibnamefont
  {Palma}},\ }\bibfield  {title} {\enquote {\bibinfo {title} {Photon
  localization versus population trapping in a coupled-cavity array},}\ }\href
  {\doibase 10.1103/PhysRevA.89.053826} {\bibfield  {journal} {\bibinfo
  {journal} {Phys. Rev. A}\ }\textbf {\bibinfo {volume} {89}},\ \bibinfo
  {pages} {053826} (\bibinfo {year} {2014})}\BibitemShut {NoStop}%
\bibitem [{\citenamefont {Calaj\'o}\ \emph {et~al.}(2016)\citenamefont
  {Calaj\'o}, \citenamefont {Ciccarello}, \citenamefont {Chang},\ and\
  \citenamefont {Rabl}}]{CalajoPRA16}%
  \BibitemOpen
  \bibfield  {author} {\bibinfo {author} {\bibfnamefont {G.}~\bibnamefont
  {Calaj\'o}}, \bibinfo {author} {\bibfnamefont {F.}~\bibnamefont
  {Ciccarello}}, \bibinfo {author} {\bibfnamefont {D.~E.}\ \bibnamefont
  {Chang}}, \ and\ \bibinfo {author} {\bibfnamefont {P.}~\bibnamefont {Rabl}},\
  }\bibfield  {title} {\enquote {\bibinfo {title} {Atom-field dressed states in
  slow-light waveguide {QED}},}\ }\href {\doibase 10.1103/PhysRevA.93.033833}
  {\bibfield  {journal} {\bibinfo  {journal} {Phys. Rev. A}\ }\textbf {\bibinfo
  {volume} {93}},\ \bibinfo {pages} {033833} (\bibinfo {year}
  {2016})}\BibitemShut {NoStop}%
\bibitem [{\citenamefont {Shi}\ \emph {et~al.}(2016)\citenamefont {Shi},
  \citenamefont {Wu}, \citenamefont {Gonz\'alez-Tudela},\ and\ \citenamefont
  {Cirac}}]{ShiPRX16}%
  \BibitemOpen
  \bibfield  {author} {\bibinfo {author} {\bibfnamefont {T.}~\bibnamefont
  {Shi}}, \bibinfo {author} {\bibfnamefont {Y.-H.}\ \bibnamefont {Wu}},
  \bibinfo {author} {\bibfnamefont {A.}~\bibnamefont {Gonz\'alez-Tudela}}, \
  and\ \bibinfo {author} {\bibfnamefont {J.~I.}\ \bibnamefont {Cirac}},\
  }\bibfield  {title} {\enquote {\bibinfo {title} {Bound states in boson
  impurity models},}\ }\href {\doibase 10.1103/PhysRevX.6.021027} {\bibfield
  {journal} {\bibinfo  {journal} {Phys. Rev. X}\ }\textbf {\bibinfo {volume}
  {6}},\ \bibinfo {pages} {021027} (\bibinfo {year} {2016})}\BibitemShut
  {NoStop}%
\bibitem [{\citenamefont {Kocaba{\c s}}(2016)}]{KocabasPRA16}%
  \BibitemOpen
  \bibfield  {author} {\bibinfo {author} {\bibfnamefont {{\c S}.~E.}\
  \bibnamefont {Kocaba{\c s}}},\ }\bibfield  {title} {\enquote {\bibinfo
  {title} {{Effects of modal dispersion on few-photon--qubit scattering in
  one-dimensional waveguides}},}\ }\href {\doibase 10.1103/PhysRevA.93.033829}
  {\bibfield  {journal} {\bibinfo  {journal} {Phys. Rev. A}\ }\textbf {\bibinfo
  {volume} {93}},\ \bibinfo {pages} {033829} (\bibinfo {year}
  {2016})}\BibitemShut {NoStop}%
\bibitem [{\citenamefont {Shukla}\ and\ \citenamefont
  {Eliasson}(2011)}]{ShuklaQPlasmasRMP11}%
  \BibitemOpen
  \bibfield  {author} {\bibinfo {author} {\bibfnamefont {P.~K.}\ \bibnamefont
  {Shukla}}\ and\ \bibinfo {author} {\bibfnamefont {B.}~\bibnamefont
  {Eliasson}},\ }\bibfield  {title} {\enquote {\bibinfo {title} {Colloquium:
  {Nonlinear} collective interactions in quantum plasmas with degenerate
  electron fluids},}\ }\href {\doibase 10.1103/RevModPhys.83.885} {\bibfield
  {journal} {\bibinfo  {journal} {Rev. Mod. Phys.}\ }\textbf {\bibinfo {volume}
  {83}},\ \bibinfo {pages} {885--906} (\bibinfo {year} {2011})}\BibitemShut
  {NoStop}%
\bibitem [{\citenamefont {Garc\'ia-Mata}\ \emph {et~al.}(2014)\citenamefont
  {Garc\'ia-Mata}, \citenamefont {Pineda},\ and\ \citenamefont
  {Wisniacki}}]{GarciaMataJPA14}%
  \BibitemOpen
  \bibfield  {author} {\bibinfo {author} {\bibfnamefont {I.}~\bibnamefont
  {Garc\'ia-Mata}}, \bibinfo {author} {\bibfnamefont {C.}~\bibnamefont
  {Pineda}}, \ and\ \bibinfo {author} {\bibfnamefont {D.~A.}\ \bibnamefont
  {Wisniacki}},\ }\bibfield  {title} {\enquote {\bibinfo {title} {Quantum
  non-{Markovian} behavior at the chaos border},}\ }\href {\doibase
  10.1088/1751-8113/47/11/115301} {\bibfield  {journal} {\bibinfo  {journal}
  {J. Phys. A}\ }\textbf {\bibinfo {volume} {47}},\ \bibinfo {pages} {115301}
  (\bibinfo {year} {2014})}\BibitemShut {NoStop}%
\bibitem [{\citenamefont {Swingle}(2018)}]{SwingleOtocNPhys18}%
  \BibitemOpen
  \bibfield  {author} {\bibinfo {author} {\bibfnamefont {B.}~\bibnamefont
  {Swingle}},\ }\bibfield  {title} {\enquote {\bibinfo {title} {Unscrambling
  the physics of out-of-time-order correlators},}\ }\href {\doibase
  10.1038/s41567-018-0295-5} {\bibfield  {journal} {\bibinfo  {journal} {Nat.
  Phys.}\ }\textbf {\bibinfo {volume} {14}},\ \bibinfo {pages} {988--990}
  (\bibinfo {year} {2018})}\BibitemShut {NoStop}%
\bibitem [{\citenamefont {Ritsch}\ \emph {et~al.}(2013)\citenamefont {Ritsch},
  \citenamefont {Domokos}, \citenamefont {Brennecke},\ and\ \citenamefont
  {Esslinger}}]{RitschRMP13}%
  \BibitemOpen
  \bibfield  {author} {\bibinfo {author} {\bibfnamefont {H.}~\bibnamefont
  {Ritsch}}, \bibinfo {author} {\bibfnamefont {P.}~\bibnamefont {Domokos}},
  \bibinfo {author} {\bibfnamefont {F.}~\bibnamefont {Brennecke}}, \ and\
  \bibinfo {author} {\bibfnamefont {T.}~\bibnamefont {Esslinger}},\ }\bibfield
  {title} {\enquote {\bibinfo {title} {Cold atoms in cavity-generated dynamical
  optical potentials},}\ }\href {\doibase 10.1103/RevModPhys.85.553} {\bibfield
   {journal} {\bibinfo  {journal} {Rev. Mod. Phys.}\ }\textbf {\bibinfo
  {volume} {85}},\ \bibinfo {pages} {553--601} (\bibinfo {year}
  {2013})}\BibitemShut {NoStop}%
\bibitem [{\citenamefont {Aspelmeyer}\ \emph {et~al.}(2014)\citenamefont
  {Aspelmeyer}, \citenamefont {Kippenberg},\ and\ \citenamefont
  {Marquardt}}]{AspelmeyerOptomechRMP14}%
  \BibitemOpen
  \bibfield  {author} {\bibinfo {author} {\bibfnamefont {M.}~\bibnamefont
  {Aspelmeyer}}, \bibinfo {author} {\bibfnamefont {T.~J.}\ \bibnamefont
  {Kippenberg}}, \ and\ \bibinfo {author} {\bibfnamefont {F.}~\bibnamefont
  {Marquardt}},\ }\bibfield  {title} {\enquote {\bibinfo {title} {Cavity
  optomechanics},}\ }\href {\doibase 10.1103/RevModPhys.86.1391} {\bibfield
  {journal} {\bibinfo  {journal} {Rev. Mod. Phys.}\ }\textbf {\bibinfo {volume}
  {86}},\ \bibinfo {pages} {1391--1452} (\bibinfo {year} {2014})}\BibitemShut
  {NoStop}%
\bibitem [{\citenamefont {Wang}\ and\ \citenamefont
  {Safavi-Naeini}(2017)}]{WangNatComm17}%
  \BibitemOpen
  \bibfield  {author} {\bibinfo {author} {\bibfnamefont {Z.}~\bibnamefont
  {Wang}}\ and\ \bibinfo {author} {\bibfnamefont {A.~H.}\ \bibnamefont
  {Safavi-Naeini}},\ }\bibfield  {title} {\enquote {\bibinfo {title} {Enhancing
  a slow and weak optomechanical nonlinearity with delayed quantum feedback},}\
  }\href {\doibase 10.1038/ncomms15886} {\bibfield  {journal} {\bibinfo
  {journal} {Nat. Commun.}\ }\textbf {\bibinfo {volume} {8}},\ \bibinfo {pages}
  {15886} (\bibinfo {year} {2017})}\BibitemShut {NoStop}%
\bibitem [{\citenamefont {Rossi}\ \emph {et~al.}(2018)\citenamefont {Rossi},
  \citenamefont {Kralj}, \citenamefont {Zippilli}, \citenamefont {Natali},
  \citenamefont {Borrielli}, \citenamefont {Pandraud}, \citenamefont {Serra},
  \citenamefont {Di~Giuseppe},\ and\ \citenamefont {Vitali}}]{RossiPRL18}%
  \BibitemOpen
  \bibfield  {author} {\bibinfo {author} {\bibfnamefont {M.}~\bibnamefont
  {Rossi}}, \bibinfo {author} {\bibfnamefont {N.}~\bibnamefont {Kralj}},
  \bibinfo {author} {\bibfnamefont {S.}~\bibnamefont {Zippilli}}, \bibinfo
  {author} {\bibfnamefont {R.}~\bibnamefont {Natali}}, \bibinfo {author}
  {\bibfnamefont {A.}~\bibnamefont {Borrielli}}, \bibinfo {author}
  {\bibfnamefont {G.}~\bibnamefont {Pandraud}}, \bibinfo {author}
  {\bibfnamefont {E.}~\bibnamefont {Serra}}, \bibinfo {author} {\bibfnamefont
  {G.}~\bibnamefont {Di~Giuseppe}}, \ and\ \bibinfo {author} {\bibfnamefont
  {D.}~\bibnamefont {Vitali}},\ }\bibfield  {title} {\enquote {\bibinfo {title}
  {Normal-mode splitting in a weakly coupled optomechanical system},}\ }\href
  {\doibase 10.1103/PhysRevLett.120.073601} {\bibfield  {journal} {\bibinfo
  {journal} {Phys. Rev. Lett.}\ }\textbf {\bibinfo {volume} {120}},\ \bibinfo
  {pages} {073601} (\bibinfo {year} {2018})}\BibitemShut {NoStop}%
\bibitem [{\citenamefont {Shen}\ and\ \citenamefont {Fan}(2007)}]{ShenPRA07}%
  \BibitemOpen
  \bibfield  {author} {\bibinfo {author} {\bibfnamefont {J.-T.}\ \bibnamefont
  {Shen}}\ and\ \bibinfo {author} {\bibfnamefont {S.}~\bibnamefont {Fan}},\
  }\bibfield  {title} {\enquote {\bibinfo {title} {Strongly correlated
  multiparticle transport in one dimension through a quantum impurity},}\
  }\href {\doibase 10.1103/PhysRevA.76.062709} {\bibfield  {journal} {\bibinfo
  {journal} {Phys. Rev. A}\ }\textbf {\bibinfo {volume} {76}},\ \bibinfo
  {pages} {062709} (\bibinfo {year} {2007})}\BibitemShut {NoStop}%
\bibitem [{\citenamefont {Shen}\ and\ \citenamefont {Fan}(2009)}]{ShenPRA09I}%
  \BibitemOpen
  \bibfield  {author} {\bibinfo {author} {\bibfnamefont {J.-T.}\ \bibnamefont
  {Shen}}\ and\ \bibinfo {author} {\bibfnamefont {S.}~\bibnamefont {Fan}},\
  }\bibfield  {title} {\enquote {\bibinfo {title} {{Theory of single-photon
  transport in a single-mode waveguide}},}\ }\href {\doibase
  10.1103/PhysRevA.79.023837} {\bibfield  {journal} {\bibinfo  {journal} {Phys.
  Rev. A}\ }\textbf {\bibinfo {volume} {79}},\ \bibinfo {pages} {023837}
  (\bibinfo {year} {2009})}\BibitemShut {NoStop}%
\bibitem [{\citenamefont {Zheng}\ \emph {et~al.}(2010)\citenamefont {Zheng},
  \citenamefont {Gauthier},\ and\ \citenamefont {Baranger}}]{ZhengPRA10}%
  \BibitemOpen
  \bibfield  {author} {\bibinfo {author} {\bibfnamefont {H.}~\bibnamefont
  {Zheng}}, \bibinfo {author} {\bibfnamefont {D.~J.}\ \bibnamefont {Gauthier}},
  \ and\ \bibinfo {author} {\bibfnamefont {H.~U.}\ \bibnamefont {Baranger}},\
  }\bibfield  {title} {\enquote {\bibinfo {title} {{Waveguide {QED}: Many-body
  bound-state effects in coherent and {F}ock-state scattering from a two-level
  system}},}\ }\href {\doibase 10.1103/PhysRevA.82.063816} {\bibfield
  {journal} {\bibinfo  {journal} {Phys. Rev. A}\ }\textbf {\bibinfo {volume}
  {82}},\ \bibinfo {pages} {063816} (\bibinfo {year} {2010})}\BibitemShut
  {NoStop}%
\bibitem [{\citenamefont {Zheng}\ and\ \citenamefont
  {Baranger}(2013)}]{ZhengPRL13}%
  \BibitemOpen
  \bibfield  {author} {\bibinfo {author} {\bibfnamefont {H.}~\bibnamefont
  {Zheng}}\ and\ \bibinfo {author} {\bibfnamefont {H.~U.}\ \bibnamefont
  {Baranger}},\ }\bibfield  {title} {\enquote {\bibinfo {title} {Persistent
  quantum beats and long-distance entanglement from waveguide-mediated
  interactions},}\ }\href {\doibase 10.1103/PhysRevLett.110.113601} {\bibfield
  {journal} {\bibinfo  {journal} {Phys. Rev. Lett.}\ }\textbf {\bibinfo
  {volume} {110}},\ \bibinfo {pages} {113601} (\bibinfo {year}
  {2013})}\BibitemShut {NoStop}%
\bibitem [{not()}]{note}%
  \BibitemOpen
  \href@noop {} {}\bibinfo {note} {Here the plus sign occurs for $m$ even,
  while the minus sign is for $m$ odd.}\BibitemShut {Stop}%
\bibitem [{\citenamefont {Tufarelli}\ \emph {et~al.}(2014)\citenamefont
  {Tufarelli}, \citenamefont {Kim},\ and\ \citenamefont
  {Ciccarello}}]{TufarelliPRA14}%
  \BibitemOpen
  \bibfield  {author} {\bibinfo {author} {\bibfnamefont {T.}~\bibnamefont
  {Tufarelli}}, \bibinfo {author} {\bibfnamefont {M.~S.}\ \bibnamefont {Kim}},
  \ and\ \bibinfo {author} {\bibfnamefont {F.}~\bibnamefont {Ciccarello}},\
  }\bibfield  {title} {\enquote {\bibinfo {title} {{Non-Markovianity of a
  quantum emitter in front of a mirror}},}\ }\href {\doibase
  10.1103/PhysRevA.90.012113} {\bibfield  {journal} {\bibinfo  {journal} {Phys.
  Rev. A}\ }\textbf {\bibinfo {volume} {90}},\ \bibinfo {pages} {012113}
  (\bibinfo {year} {2014})}\BibitemShut {NoStop}%
\bibitem [{\citenamefont {Hoi}\ \emph {et~al.}(2015)\citenamefont {Hoi},
  \citenamefont {Kockum}, \citenamefont {Tornberg}, \citenamefont
  {Pourkabirian}, \citenamefont {Johansson}, \citenamefont {Delsing},\ and\
  \citenamefont {Wilson}}]{HoiNatPhy15}%
  \BibitemOpen
  \bibfield  {author} {\bibinfo {author} {\bibfnamefont {I.-C.}\ \bibnamefont
  {Hoi}}, \bibinfo {author} {\bibfnamefont {A.~F.}\ \bibnamefont {Kockum}},
  \bibinfo {author} {\bibfnamefont {L.}~\bibnamefont {Tornberg}}, \bibinfo
  {author} {\bibfnamefont {A.}~\bibnamefont {Pourkabirian}}, \bibinfo {author}
  {\bibfnamefont {G.}~\bibnamefont {Johansson}}, \bibinfo {author}
  {\bibfnamefont {P.}~\bibnamefont {Delsing}}, \ and\ \bibinfo {author}
  {\bibfnamefont {C.~M.}\ \bibnamefont {Wilson}},\ }\bibfield  {title}
  {\enquote {\bibinfo {title} {{Probing the quantum vacuum with an artificial
  atom in front of a mirror}},}\ }\href {\doibase 10.1038/nphys3484} {\bibfield
   {journal} {\bibinfo  {journal} {Nat. Phys.}\ }\textbf {\bibinfo {volume}
  {11}},\ \bibinfo {pages} {1045--1049} (\bibinfo {year} {2015})}\BibitemShut
  {NoStop}%
\bibitem [{Sup()}]{SupMat}%
  \BibitemOpen
  \href@noop {} {}\bibinfo {note} {See Supplementary Material at [URL] for
  computational methods used in this work (see also Ref. \cite{FangCPC18}),
  derivation of \eq\eqref{Ps}, optimal $\Delta k $ used in
  \fig\figurepanel{Fig3_optimization}{a}, explanation of the method used to
  increase BIC generation probability in \fig\ref{Fig5_BIC}, additional notes
  on the two-qubit BIC, impact of nonzero loss to photon trapping, and the
  importance of having nonlinear emitters.}\BibitemShut {Stop}%
\bibitem [{\citenamefont {Stobi\'{n}ska}\ \emph {et~al.}(2009)\citenamefont
  {Stobi\'{n}ska}, \citenamefont {Alber},\ and\ \citenamefont
  {Leuchs}}]{StobinskaEPL09}%
  \BibitemOpen
  \bibfield  {author} {\bibinfo {author} {\bibfnamefont {M.}~\bibnamefont
  {Stobi\'{n}ska}}, \bibinfo {author} {\bibfnamefont {G.}~\bibnamefont
  {Alber}}, \ and\ \bibinfo {author} {\bibfnamefont {G.}~\bibnamefont
  {Leuchs}},\ }\bibfield  {title} {\enquote {\bibinfo {title} {Perfect
  excitation of a matter qubit by a single photon in free space},}\ }\href
  {\doibase 10.1209/0295-5075/86/14007} {\bibfield  {journal} {\bibinfo
  {journal} {Eur.ophys. Lett.}\ }\textbf {\bibinfo {volume} {86}},\ \bibinfo
  {pages} {14007} (\bibinfo {year} {2009})}\BibitemShut {NoStop}%
\bibitem [{\citenamefont {Baragiola}\ \emph {et~al.}(2012)\citenamefont
  {Baragiola}, \citenamefont {Cook}, \citenamefont {Bra{\'n}czyk},\ and\
  \citenamefont {Combes}}]{BaragiolaPRA12}%
  \BibitemOpen
  \bibfield  {author} {\bibinfo {author} {\bibfnamefont {B.~Q.}\ \bibnamefont
  {Baragiola}}, \bibinfo {author} {\bibfnamefont {R.~L.}\ \bibnamefont {Cook}},
  \bibinfo {author} {\bibfnamefont {A.~M.}\ \bibnamefont {Bra{\'n}czyk}}, \
  and\ \bibinfo {author} {\bibfnamefont {J.}~\bibnamefont {Combes}},\
  }\bibfield  {title} {\enquote {\bibinfo {title} {{N-photon wave packets
  interacting with an arbitrary quantum system}},}\ }\href {\doibase
  10.1103/PhysRevA.86.013811} {\bibfield  {journal} {\bibinfo  {journal} {Phys.
  Rev. A}\ }\textbf {\bibinfo {volume} {86}},\ \bibinfo {pages} {013811}
  (\bibinfo {year} {2012})}\BibitemShut {NoStop}%
\bibitem [{\citenamefont {Fang}\ \emph {et~al.}(2018)\citenamefont {Fang},
  \citenamefont {Ciccarello},\ and\ \citenamefont {Baranger}}]{FangNJP18}%
  \BibitemOpen
  \bibfield  {author} {\bibinfo {author} {\bibfnamefont {Y.-L.~L.}\
  \bibnamefont {Fang}}, \bibinfo {author} {\bibfnamefont {F.}~\bibnamefont
  {Ciccarello}}, \ and\ \bibinfo {author} {\bibfnamefont {H.~U.}\ \bibnamefont
  {Baranger}},\ }\bibfield  {title} {\enquote {\bibinfo {title} {{Non-Markovian
  dynamics of a qubit due to single-photon scattering in a waveguide}},}\
  }\href {\doibase 10.1088/1367-2630/aaba5d} {\bibfield  {journal} {\bibinfo
  {journal} {New J. Phys.}\ }\textbf {\bibinfo {volume} {20}},\ \bibinfo
  {pages} {043035} (\bibinfo {year} {2018})}\BibitemShut {NoStop}%
\bibitem [{\citenamefont {Fang}\ and\ \citenamefont
  {Baranger}(2016)}]{FangPE16}%
  \BibitemOpen
  \bibfield  {author} {\bibinfo {author} {\bibfnamefont {Y.-L.~L.}\
  \bibnamefont {Fang}}\ and\ \bibinfo {author} {\bibfnamefont {H.}~\bibnamefont
  {Baranger}},\ }\bibfield  {title} {\enquote {\bibinfo {title} {{Photon
  correlations generated by inelastic scattering in a one-dimensional waveguide
  coupled to three-level systems}},}\ }\href {\doibase
  10.1016/j.physe.2015.11.004} {\bibfield  {journal} {\bibinfo  {journal}
  {Physica}\ }\textbf {\bibinfo {volume} {78E}},\ \bibinfo {pages} {92--99}
  (\bibinfo {year} {2016})}\BibitemShut {NoStop}%
\bibitem [{\citenamefont {Fang}\ and\ \citenamefont
  {Baranger}(2017)}]{FangPRA17}%
  \BibitemOpen
  \bibfield  {author} {\bibinfo {author} {\bibfnamefont {Y.-L.~L.}\
  \bibnamefont {Fang}}\ and\ \bibinfo {author} {\bibfnamefont {H.~U.}\
  \bibnamefont {Baranger}},\ }\bibfield  {title} {\enquote {\bibinfo {title}
  {{Multiple emitters in a waveguide: Nonreciprocity and correlated photons at
  perfect elastic transmission}},}\ }\href {\doibase
  10.1103/PhysRevA.96.013842} {\bibfield  {journal} {\bibinfo  {journal} {Phys.
  Rev. A}\ }\textbf {\bibinfo {volume} {96}},\ \bibinfo {pages} {013842}
  (\bibinfo {year} {2017})}\BibitemShut {NoStop}%
\bibitem [{\citenamefont {Rephaeli}\ and\ \citenamefont
  {Fan}(2012)}]{RephaeliPRL12}%
  \BibitemOpen
  \bibfield  {author} {\bibinfo {author} {\bibfnamefont {E.}~\bibnamefont
  {Rephaeli}}\ and\ \bibinfo {author} {\bibfnamefont {S.}~\bibnamefont {Fan}},\
  }\bibfield  {title} {\enquote {\bibinfo {title} {Stimulated emission from a
  single excited atom in a waveguide},}\ }\href {\doibase
  10.1103/PhysRevLett.108.143602} {\bibfield  {journal} {\bibinfo  {journal}
  {Phys. Rev. Lett.}\ }\textbf {\bibinfo {volume} {108}},\ \bibinfo {pages}
  {143602} (\bibinfo {year} {2012})}\BibitemShut {NoStop}%
\bibitem [{\citenamefont {Cotrufo}\ and\ \citenamefont
  {Al{\`u}}(2018)}]{CotrufoAluarXiv18}%
  \BibitemOpen
  \bibfield  {author} {\bibinfo {author} {\bibfnamefont {M.}~\bibnamefont
  {Cotrufo}}\ and\ \bibinfo {author} {\bibfnamefont {A.}~\bibnamefont
  {Al{\`u}}},\ }\href@noop {} {\enquote {\bibinfo {title} {Single-photon
  embedded eigenstates in coupled cavity-atom systems},}\ } (\bibinfo {year}
  {2018}),\ \Eprint {http://arxiv.org/abs/1805.03287} {arXiv:1805.03287}
  \BibitemShut {NoStop}%
\bibitem [{\citenamefont {Lalumi\`ere}\ \emph {et~al.}(2013)\citenamefont
  {Lalumi\`ere}, \citenamefont {Sanders}, \citenamefont {van Loo},
  \citenamefont {Fedorov}, \citenamefont {Wallraff},\ and\ \citenamefont
  {Blais}}]{LalumierePRA13}%
  \BibitemOpen
  \bibfield  {author} {\bibinfo {author} {\bibfnamefont {K.}~\bibnamefont
  {Lalumi\`ere}}, \bibinfo {author} {\bibfnamefont {B.~C.}\ \bibnamefont
  {Sanders}}, \bibinfo {author} {\bibfnamefont {A.~F.}\ \bibnamefont {van
  Loo}}, \bibinfo {author} {\bibfnamefont {A.}~\bibnamefont {Fedorov}},
  \bibinfo {author} {\bibfnamefont {A.}~\bibnamefont {Wallraff}}, \ and\
  \bibinfo {author} {\bibfnamefont {A.}~\bibnamefont {Blais}},\ }\bibfield
  {title} {\enquote {\bibinfo {title} {Input-output theory for waveguide {QED}
  with an ensemble of inhomogeneous atoms},}\ }\href {\doibase
  10.1103/PhysRevA.88.043806} {\bibfield  {journal} {\bibinfo  {journal} {Phys.
  Rev. A}\ }\textbf {\bibinfo {volume} {88}},\ \bibinfo {pages} {043806}
  (\bibinfo {year} {2013})}\BibitemShut {NoStop}%
\bibitem [{\citenamefont {van Loo}\ \emph {et~al.}(2013)\citenamefont {van
  Loo}, \citenamefont {Fedorov}, \citenamefont {Lalumi\`ere}, \citenamefont
  {Sanders}, \citenamefont {Blais},\ and\ \citenamefont
  {Wallraff}}]{vanLooScience13}%
  \BibitemOpen
  \bibfield  {author} {\bibinfo {author} {\bibfnamefont {A.~F.}\ \bibnamefont
  {van Loo}}, \bibinfo {author} {\bibfnamefont {A.}~\bibnamefont {Fedorov}},
  \bibinfo {author} {\bibfnamefont {K.}~\bibnamefont {Lalumi\`ere}}, \bibinfo
  {author} {\bibfnamefont {B.~C.}\ \bibnamefont {Sanders}}, \bibinfo {author}
  {\bibfnamefont {A.}~\bibnamefont {Blais}}, \ and\ \bibinfo {author}
  {\bibfnamefont {A.}~\bibnamefont {Wallraff}},\ }\bibfield  {title} {\enquote
  {\bibinfo {title} {{Photon-Mediated Interactions Between Distant Artificial
  Atoms}},}\ }\href {\doibase 10.1126/science.1244324} {\bibfield  {journal}
  {\bibinfo  {journal} {Science}\ }\textbf {\bibinfo {volume} {342}},\ \bibinfo
  {pages} {1494--1496} (\bibinfo {year} {2013})}\BibitemShut {NoStop}%
\bibitem [{\citenamefont {Fang}\ and\ \citenamefont
  {Baranger}(2015)}]{FangPRA15}%
  \BibitemOpen
  \bibfield  {author} {\bibinfo {author} {\bibfnamefont {Y.-L.~L.}\
  \bibnamefont {Fang}}\ and\ \bibinfo {author} {\bibfnamefont {H.~U.}\
  \bibnamefont {Baranger}},\ }\bibfield  {title} {\enquote {\bibinfo {title}
  {{Waveguide QED: Power spectra and correlations of two photons scattered off
  multiple distant qubits and a mirror}},}\ }\href {\doibase
  10.1103/PhysRevA.91.053845} {\bibfield  {journal} {\bibinfo  {journal} {Phys.
  Rev. A}\ }\textbf {\bibinfo {volume} {91}},\ \bibinfo {pages} {053845}
  (\bibinfo {year} {2015})},\ \bibinfo {note}
  {\href{http://link.aps.org/doi/10.1103/PhysRevA.96.059904}{\textit{ibid.}
  \textbf{96}, 059904(E) (2017).}}\BibitemShut {Stop}%
\bibitem [{\citenamefont {Wootters}(1998)}]{WoottersPRL98}%
  \BibitemOpen
  \bibfield  {author} {\bibinfo {author} {\bibfnamefont {W.~K.}\ \bibnamefont
  {Wootters}},\ }\bibfield  {title} {\enquote {\bibinfo {title} {Entanglement
  of formation of an arbitrary state of two qubits},}\ }\href {\doibase
  10.1103/PhysRevLett.80.2245} {\bibfield  {journal} {\bibinfo  {journal}
  {Phys. Rev. Lett.}\ }\textbf {\bibinfo {volume} {80}},\ \bibinfo {pages}
  {2245--2248} (\bibinfo {year} {1998})}\BibitemShut {NoStop}%
\bibitem [{\citenamefont {Ficek}\ and\ \citenamefont
  {Tana\ifmmode~\acute{s}\else \'{s}\fi{}}(2008)}]{FicekPRA08}%
  \BibitemOpen
  \bibfield  {author} {\bibinfo {author} {\bibfnamefont {Z.}~\bibnamefont
  {Ficek}}\ and\ \bibinfo {author} {\bibfnamefont {R.}~\bibnamefont
  {Tana\ifmmode~\acute{s}\else \'{s}\fi{}}},\ }\bibfield  {title} {\enquote
  {\bibinfo {title} {Delayed sudden birth of entanglement},}\ }\href {\doibase
  10.1103/PhysRevA.77.054301} {\bibfield  {journal} {\bibinfo  {journal} {Phys.
  Rev. A}\ }\textbf {\bibinfo {volume} {77}},\ \bibinfo {pages} {054301}
  (\bibinfo {year} {2008})}\BibitemShut {NoStop}%
\bibitem [{\citenamefont {Dorner}\ and\ \citenamefont
  {Zoller}(2002)}]{DornerPRA02}%
  \BibitemOpen
  \bibfield  {author} {\bibinfo {author} {\bibfnamefont {U.}~\bibnamefont
  {Dorner}}\ and\ \bibinfo {author} {\bibfnamefont {P.}~\bibnamefont
  {Zoller}},\ }\bibfield  {title} {\enquote {\bibinfo {title} {{Laser-driven
  atoms in half-cavities}},}\ }\href {\doibase 10.1103/PhysRevA.66.023816}
  {\bibfield  {journal} {\bibinfo  {journal} {Phys. Rev. A}\ }\textbf {\bibinfo
  {volume} {66}},\ \bibinfo {pages} {023816} (\bibinfo {year}
  {2002})}\BibitemShut {NoStop}%
\bibitem [{\citenamefont {Carmele}\ \emph {et~al.}(2013)\citenamefont
  {Carmele}, \citenamefont {Kabuss}, \citenamefont {Schulze}, \citenamefont
  {Reitzenstein},\ and\ \citenamefont {Knorr}}]{CarmelePRL13}%
  \BibitemOpen
  \bibfield  {author} {\bibinfo {author} {\bibfnamefont {A.}~\bibnamefont
  {Carmele}}, \bibinfo {author} {\bibfnamefont {J.}~\bibnamefont {Kabuss}},
  \bibinfo {author} {\bibfnamefont {F.}~\bibnamefont {Schulze}}, \bibinfo
  {author} {\bibfnamefont {S.}~\bibnamefont {Reitzenstein}}, \ and\ \bibinfo
  {author} {\bibfnamefont {A.}~\bibnamefont {Knorr}},\ }\bibfield  {title}
  {\enquote {\bibinfo {title} {{Single Photon Delayed Feedback: A Way to
  Stabilize Intrinsic Quantum Cavity Electrodynamics}},}\ }\href {\doibase
  10.1103/PhysRevLett.110.013601} {\bibfield  {journal} {\bibinfo  {journal}
  {Phys. Rev. Lett.}\ }\textbf {\bibinfo {volume} {110}},\ \bibinfo {pages}
  {013601} (\bibinfo {year} {2013})}\BibitemShut {NoStop}%
\bibitem [{\citenamefont {Laakso}\ and\ \citenamefont
  {Pletyukhov}(2014)}]{LaaksoPRL14}%
  \BibitemOpen
  \bibfield  {author} {\bibinfo {author} {\bibfnamefont {M.}~\bibnamefont
  {Laakso}}\ and\ \bibinfo {author} {\bibfnamefont {M.}~\bibnamefont
  {Pletyukhov}},\ }\bibfield  {title} {\enquote {\bibinfo {title} {Scattering
  of two photons from two distant qubits: Exact solution},}\ }\href {\doibase
  10.1103/PhysRevLett.113.183601} {\bibfield  {journal} {\bibinfo  {journal}
  {Phys. Rev. Lett.}\ }\textbf {\bibinfo {volume} {113}},\ \bibinfo {pages}
  {183601} (\bibinfo {year} {2014})}\BibitemShut {NoStop}%
\bibitem [{\citenamefont {Grimsmo}(2015)}]{GrimsmoPRL15}%
  \BibitemOpen
  \bibfield  {author} {\bibinfo {author} {\bibfnamefont {A.~L.}\ \bibnamefont
  {Grimsmo}},\ }\bibfield  {title} {\enquote {\bibinfo {title} {{Time-Delayed
  Quantum Feedback Control}},}\ }\href {\doibase
  10.1103/PhysRevLett.115.060402} {\bibfield  {journal} {\bibinfo  {journal}
  {Phys. Rev. Lett.}\ }\textbf {\bibinfo {volume} {115}},\ \bibinfo {pages}
  {060402} (\bibinfo {year} {2015})}\BibitemShut {NoStop}%
\bibitem [{\citenamefont {Pichler}\ and\ \citenamefont
  {Zoller}(2016)}]{PichlerPRL16}%
  \BibitemOpen
  \bibfield  {author} {\bibinfo {author} {\bibfnamefont {H.}~\bibnamefont
  {Pichler}}\ and\ \bibinfo {author} {\bibfnamefont {P.}~\bibnamefont
  {Zoller}},\ }\bibfield  {title} {\enquote {\bibinfo {title} {{Photonic
  Circuits with Time Delays and Quantum Feedback}},}\ }\href {\doibase
  10.1103/PhysRevLett.116.093601} {\bibfield  {journal} {\bibinfo  {journal}
  {Phys. Rev. Lett.}\ }\textbf {\bibinfo {volume} {116}},\ \bibinfo {pages}
  {093601} (\bibinfo {year} {2016})}\BibitemShut {NoStop}%
\bibitem [{\citenamefont {Ramos}\ \emph {et~al.}(2016)\citenamefont {Ramos},
  \citenamefont {Vermersch}, \citenamefont {Hauke}, \citenamefont {Pichler},\
  and\ \citenamefont {Zoller}}]{RamosPRA16}%
  \BibitemOpen
  \bibfield  {author} {\bibinfo {author} {\bibfnamefont {T.}~\bibnamefont
  {Ramos}}, \bibinfo {author} {\bibfnamefont {B.}~\bibnamefont {Vermersch}},
  \bibinfo {author} {\bibfnamefont {P.}~\bibnamefont {Hauke}}, \bibinfo
  {author} {\bibfnamefont {H.}~\bibnamefont {Pichler}}, \ and\ \bibinfo
  {author} {\bibfnamefont {P.}~\bibnamefont {Zoller}},\ }\bibfield  {title}
  {\enquote {\bibinfo {title} {{Non-Markovian dynamics in chiral quantum
  networks with spins and photons}},}\ }\href {\doibase
  10.1103/PhysRevA.93.062104} {\bibfield  {journal} {\bibinfo  {journal} {Phys.
  Rev. A}\ }\textbf {\bibinfo {volume} {93}},\ \bibinfo {pages} {062104}
  (\bibinfo {year} {2016})}\BibitemShut {NoStop}%
\bibitem [{\citenamefont {Tabak}\ and\ \citenamefont
  {Mabuchi}(2016)}]{TabakEPJQT16}%
  \BibitemOpen
  \bibfield  {author} {\bibinfo {author} {\bibfnamefont {G.}~\bibnamefont
  {Tabak}}\ and\ \bibinfo {author} {\bibfnamefont {H.}~\bibnamefont
  {Mabuchi}},\ }\bibfield  {title} {\enquote {\bibinfo {title} {Trapped modes
  in linear quantum stochastic networks with delays},}\ }\href {\doibase
  10.1140/epjqt/s40507-016-0041-9} {\bibfield  {journal} {\bibinfo  {journal}
  {EPJ Quantum Technol.}\ }\textbf {\bibinfo {volume} {3}},\ \bibinfo {pages}
  {3} (\bibinfo {year} {2016})}\BibitemShut {NoStop}%
\bibitem [{\citenamefont {Whalen}\ \emph {et~al.}(2017)\citenamefont {Whalen},
  \citenamefont {Grimsmo},\ and\ \citenamefont {Carmichael}}]{WhalenQST17}%
  \BibitemOpen
  \bibfield  {author} {\bibinfo {author} {\bibfnamefont {S.~J.}\ \bibnamefont
  {Whalen}}, \bibinfo {author} {\bibfnamefont {A.~L.}\ \bibnamefont {Grimsmo}},
  \ and\ \bibinfo {author} {\bibfnamefont {H.~J.}\ \bibnamefont {Carmichael}},\
  }\bibfield  {title} {\enquote {\bibinfo {title} {Open quantum systems with
  delayed coherent feedback},}\ }\href
  {http://stacks.iop.org/2058-9565/2/i=4/a=044008} {\bibfield  {journal}
  {\bibinfo  {journal} {Quantum Science and Technology}\ }\textbf {\bibinfo
  {volume} {2}},\ \bibinfo {pages} {044008} (\bibinfo {year}
  {2017})}\BibitemShut {NoStop}%
\bibitem [{\citenamefont {Guimond}\ \emph {et~al.}(2017)\citenamefont
  {Guimond}, \citenamefont {Pletyukhov}, \citenamefont {Pichler},\ and\
  \citenamefont {Zoller}}]{GuimondQST17}%
  \BibitemOpen
  \bibfield  {author} {\bibinfo {author} {\bibfnamefont {P.-O.}\ \bibnamefont
  {Guimond}}, \bibinfo {author} {\bibfnamefont {M.}~\bibnamefont {Pletyukhov}},
  \bibinfo {author} {\bibfnamefont {H.}~\bibnamefont {Pichler}}, \ and\
  \bibinfo {author} {\bibfnamefont {P.}~\bibnamefont {Zoller}},\ }\bibfield
  {title} {\enquote {\bibinfo {title} {Delayed coherent quantum feedback from a
  scattering theory and a matrix product state perspective},}\ }\href
  {http://stacks.iop.org/2058-9565/2/i=4/a=044012} {\bibfield  {journal}
  {\bibinfo  {journal} {Quantum Sci. Technol.}\ }\textbf {\bibinfo {volume}
  {2}},\ \bibinfo {pages} {044012} (\bibinfo {year} {2017})}\BibitemShut
  {NoStop}%
\bibitem [{\citenamefont {Pichler}\ \emph {et~al.}(2017)\citenamefont
  {Pichler}, \citenamefont {Choi}, \citenamefont {Zoller},\ and\ \citenamefont
  {Lukin}}]{PichlerPNAS17}%
  \BibitemOpen
  \bibfield  {author} {\bibinfo {author} {\bibfnamefont {H.}~\bibnamefont
  {Pichler}}, \bibinfo {author} {\bibfnamefont {S.}~\bibnamefont {Choi}},
  \bibinfo {author} {\bibfnamefont {P.}~\bibnamefont {Zoller}}, \ and\ \bibinfo
  {author} {\bibfnamefont {M.~D.}\ \bibnamefont {Lukin}},\ }\bibfield  {title}
  {\enquote {\bibinfo {title} {Universal photonic quantum computation via
  time-delayed feedback},}\ }\href {\doibase 10.1073/pnas.1711003114}
  {\bibfield  {journal} {\bibinfo  {journal} {Proc. Natl. Acad. Sci.}\ }\textbf
  {\bibinfo {volume} {114}},\ \bibinfo {pages} {11362--11367} (\bibinfo {year}
  {2017})}\BibitemShut {NoStop}%
\bibitem [{\citenamefont {Guo}\ \emph {et~al.}(2017)\citenamefont {Guo},
  \citenamefont {Grimsmo}, \citenamefont {Kockum}, \citenamefont {Pletyukhov},\
  and\ \citenamefont {Johansson}}]{GuoPRA17}%
  \BibitemOpen
  \bibfield  {author} {\bibinfo {author} {\bibfnamefont {L.}~\bibnamefont
  {Guo}}, \bibinfo {author} {\bibfnamefont {A.}~\bibnamefont {Grimsmo}},
  \bibinfo {author} {\bibfnamefont {A.~F.}\ \bibnamefont {Kockum}}, \bibinfo
  {author} {\bibfnamefont {M.}~\bibnamefont {Pletyukhov}}, \ and\ \bibinfo
  {author} {\bibfnamefont {G.}~\bibnamefont {Johansson}},\ }\bibfield  {title}
  {\enquote {\bibinfo {title} {Giant acoustic atom: A single quantum system
  with a deterministic time delay},}\ }\href {\doibase
  10.1103/PhysRevA.95.053821} {\bibfield  {journal} {\bibinfo  {journal} {Phys.
  Rev. A}\ }\textbf {\bibinfo {volume} {95}},\ \bibinfo {pages} {053821}
  (\bibinfo {year} {2017})}\BibitemShut {NoStop}%
\bibitem [{\citenamefont {Chalabi}\ and\ \citenamefont
  {Waks}(2018)}]{ChalabiPRA18}%
  \BibitemOpen
  \bibfield  {author} {\bibinfo {author} {\bibfnamefont {H.}~\bibnamefont
  {Chalabi}}\ and\ \bibinfo {author} {\bibfnamefont {E.}~\bibnamefont {Waks}},\
  }\bibfield  {title} {\enquote {\bibinfo {title} {Interaction of photons with
  a coupled atom-cavity system through a bidirectional time-delayed
  feedback},}\ }\href {\doibase 10.1103/PhysRevA.98.063832} {\bibfield
  {journal} {\bibinfo  {journal} {Phys. Rev. A}\ }\textbf {\bibinfo {volume}
  {98}},\ \bibinfo {pages} {063832} (\bibinfo {year} {2018})}\BibitemShut
  {NoStop}%
\bibitem [{\citenamefont {Tabak}\ \emph {et~al.}(2018)\citenamefont {Tabak},
  \citenamefont {Hamerly},\ and\ \citenamefont {Mabuchi}}]{TabakarXiv18}%
  \BibitemOpen
  \bibfield  {author} {\bibinfo {author} {\bibfnamefont {G.}~\bibnamefont
  {Tabak}}, \bibinfo {author} {\bibfnamefont {R.}~\bibnamefont {Hamerly}}, \
  and\ \bibinfo {author} {\bibfnamefont {H.}~\bibnamefont {Mabuchi}},\
  }\href@noop {} {\enquote {\bibinfo {title} {Factorization of linear quantum
  systems with delayed feedback},}\ } (\bibinfo {year} {2018}),\ \Eprint
  {http://arxiv.org/abs/1803.01539} {arXiv:1803.01539} \BibitemShut {NoStop}%
\bibitem [{\citenamefont {Chang}\ \emph {et~al.}(2018)\citenamefont {Chang},
  \citenamefont {Lanco}, \citenamefont {Senellart},\ and\ \citenamefont
  {Citrin}}]{ChangarXiv18}%
  \BibitemOpen
  \bibfield  {author} {\bibinfo {author} {\bibfnamefont {C.-Y.}\ \bibnamefont
  {Chang}}, \bibinfo {author} {\bibfnamefont {L.}~\bibnamefont {Lanco}},
  \bibinfo {author} {\bibfnamefont {P.}~\bibnamefont {Senellart}}, \ and\
  \bibinfo {author} {\bibfnamefont {D.~S.}\ \bibnamefont {Citrin}},\
  }\href@noop {} {\enquote {\bibinfo {title} {Quantum stabilization of a
  single-photon emitter in a coupled microcavity--half-cavity system},}\ }
  (\bibinfo {year} {2018}),\ \Eprint {http://arxiv.org/abs/1804.06734}
  {arXiv:1804.06734} \BibitemShut {NoStop}%
\bibitem [{\citenamefont {Torre}\ and\ \citenamefont
  {Illuminati}(2018)}]{TorrearXiv18}%
  \BibitemOpen
  \bibfield  {author} {\bibinfo {author} {\bibfnamefont {G.}~\bibnamefont
  {Torre}}\ and\ \bibinfo {author} {\bibfnamefont {F.}~\bibnamefont
  {Illuminati}},\ }\href@noop {} {\enquote {\bibinfo {title}
  {{Non-Markovianity-assisted optimal continuous variable quantum
  teleportation}},}\ } (\bibinfo {year} {2018}),\ \Eprint
  {http://arxiv.org/abs/1805.03617} {arXiv:1805.03617} \BibitemShut {NoStop}%
\bibitem [{\citenamefont {Andersson}\ \emph {et~al.}(2018)\citenamefont
  {Andersson}, \citenamefont {Suri}, \citenamefont {Guo}, \citenamefont
  {Aref},\ and\ \citenamefont {Delsing}}]{AnderssonarXiv18}%
  \BibitemOpen
  \bibfield  {author} {\bibinfo {author} {\bibfnamefont {G.}~\bibnamefont
  {Andersson}}, \bibinfo {author} {\bibfnamefont {B.}~\bibnamefont {Suri}},
  \bibinfo {author} {\bibfnamefont {L.}~\bibnamefont {Guo}}, \bibinfo {author}
  {\bibfnamefont {T.}~\bibnamefont {Aref}}, \ and\ \bibinfo {author}
  {\bibfnamefont {P.}~\bibnamefont {Delsing}},\ }\href@noop {} {\enquote
  {\bibinfo {title} {Nonexponential decay of a giant artificial atom},}\ }
  (\bibinfo {year} {2018}),\ \Eprint {http://arxiv.org/abs/1812.01302}
  {arXiv:1812.01302} \BibitemShut {NoStop}%
\bibitem [{\citenamefont {Fang}(2019)}]{FangCPC18}%
  \BibitemOpen
  \bibfield  {author} {\bibinfo {author} {\bibfnamefont {Y.-L.~L.}\
  \bibnamefont {Fang}},\ }\bibfield  {title} {\enquote {\bibinfo {title}
  {{FDTD: Solving 1+1D delay PDE in parallel}},}\ }\href {\doibase
  10.1016/j.cpc.2018.08.018} {\bibfield  {journal} {\bibinfo  {journal}
  {Comput. Phys. Commun.}\ }\textbf {\bibinfo {volume} {235}},\ \bibinfo
  {pages} {422 -- 432} (\bibinfo {year} {2019})}\BibitemShut {NoStop}%
\end{thebibliography}%


\begin{thebibliography}{10}%
\makeatletter
\providecommand \@ifxundefined [1]{%
 \@ifx{#1\undefined}
}%
\providecommand \@ifnum [1]{%
 \ifnum #1\expandafter \@firstoftwo
 \else \expandafter \@secondoftwo
 \fi
}%
\providecommand \@ifx [1]{%
 \ifx #1\expandafter \@firstoftwo
 \else \expandafter \@secondoftwo
 \fi
}%
\providecommand \natexlab [1]{#1}%
\providecommand \enquote  [1]{``#1''}%
\providecommand \bibnamefont  [1]{#1}%
\providecommand \bibfnamefont [1]{#1}%
\providecommand \citenamefont [1]{#1}%
\providecommand \href@noop [0]{\@secondoftwo}%
\providecommand \href [0]{\begingroup \@sanitize@url \@href}%
\providecommand \@href[1]{\@@startlink{#1}\@@href}%
\providecommand \@@href[1]{\endgroup#1\@@endlink}%
\providecommand \@sanitize@url [0]{\catcode `\\12\catcode `\$12\catcode
  `\&12\catcode `\#12\catcode `\^12\catcode `\_12\catcode `\%12\relax}%
\providecommand \@@startlink[1]{}%
\providecommand \@@endlink[0]{}%
\providecommand \url  [0]{\begingroup\@sanitize@url \@url }%
\providecommand \@url [1]{\endgroup\@href {#1}{\urlprefix }}%
\providecommand \urlprefix  [0]{URL }%
\providecommand \Eprint [0]{\href }%
\providecommand \doibase [0]{http://dx.doi.org/}%
\providecommand \selectlanguage [0]{\@gobble}%
\providecommand \bibinfo  [0]{\@secondoftwo}%
\providecommand \bibfield  [0]{\@secondoftwo}%
\providecommand \translation [1]{[#1]}%
\providecommand \BibitemOpen [0]{}%
\providecommand \bibitemStop [0]{}%
\providecommand \bibitemNoStop [0]{.\EOS\space}%
\providecommand \EOS [0]{\spacefactor3000\relax}%
\providecommand \BibitemShut  [1]{\csname bibitem#1\endcsname}%
\let\auto@bib@innerbib\@empty
%</preamble>
\bibitem [{\citenamefont {Fang}\ \emph {et~al.}(2018)\citenamefont {Fang},
  \citenamefont {Ciccarello},\ and\ \citenamefont {Baranger}}]{FangNJP18}%
  \BibitemOpen
  \bibfield  {author} {\bibinfo {author} {\bibfnamefont {Y.-L.~L.}\
  \bibnamefont {Fang}}, \bibinfo {author} {\bibfnamefont {F.}~\bibnamefont
  {Ciccarello}}, \ and\ \bibinfo {author} {\bibfnamefont {H.~U.}\ \bibnamefont
  {Baranger}},\ }\bibfield  {title} {\enquote {\bibinfo {title} {{Non-Markovian
  dynamics of a qubit due to single-photon scattering in a waveguide}},}\
  }\href {\doibase 10.1088/1367-2630/aaba5d} {\bibfield  {journal} {\bibinfo
  {journal} {New J. Phys.}\ }\textbf {\bibinfo {volume} {20}},\ \bibinfo
  {pages} {043035} (\bibinfo {year} {2018})}\BibitemShut {NoStop}%
\bibitem [{\citenamefont {Fang}(2019)}]{FangCPC18}%
  \BibitemOpen
  \bibfield  {author} {\bibinfo {author} {\bibfnamefont {Y.-L.~L.}\
  \bibnamefont {Fang}},\ }\bibfield  {title} {\enquote {\bibinfo {title}
  {{FDTD: Solving 1+1D delay PDE in parallel}},}\ }\href {\doibase
  10.1016/j.cpc.2018.08.018} {\bibfield  {journal} {\bibinfo  {journal}
  {Comput. Phys. Commun.}\ }\textbf {\bibinfo {volume} {235}},\ \bibinfo
  {pages} {422 -- 432} (\bibinfo {year} {2019})}\BibitemShut {NoStop}%
\bibitem [{\citenamefont {Calaj\'o}\ \emph {et~al.}(2016)\citenamefont
  {Calaj\'o}, \citenamefont {Ciccarello}, \citenamefont {Chang},\ and\
  \citenamefont {Rabl}}]{CalajoPRA16}%
  \BibitemOpen
  \bibfield  {author} {\bibinfo {author} {\bibfnamefont {G.}~\bibnamefont
  {Calaj\'o}}, \bibinfo {author} {\bibfnamefont {F.}~\bibnamefont
  {Ciccarello}}, \bibinfo {author} {\bibfnamefont {D.~E.}\ \bibnamefont
  {Chang}}, \ and\ \bibinfo {author} {\bibfnamefont {P.}~\bibnamefont {Rabl}},\
  }\bibfield  {title} {\enquote {\bibinfo {title} {Atom-field dressed states in
  slow-light waveguide {QED}},}\ }\href {\doibase 10.1103/PhysRevA.93.033833}
  {\bibfield  {journal} {\bibinfo  {journal} {Phys. Rev. A}\ }\textbf {\bibinfo
  {volume} {93}},\ \bibinfo {pages} {033833} (\bibinfo {year}
  {2016})}\BibitemShut {NoStop}%
\bibitem [{\citenamefont {Longo}\ \emph {et~al.}(2010)\citenamefont {Longo},
  \citenamefont {Schmitteckert},\ and\ \citenamefont {Busch}}]{LongoPRL10}%
  \BibitemOpen
  \bibfield  {author} {\bibinfo {author} {\bibfnamefont {P.}~\bibnamefont
  {Longo}}, \bibinfo {author} {\bibfnamefont {P.}~\bibnamefont
  {Schmitteckert}}, \ and\ \bibinfo {author} {\bibfnamefont {K.}~\bibnamefont
  {Busch}},\ }\bibfield  {title} {\enquote {\bibinfo {title} {Few-photon
  transport in low-dimensional systems: Interaction-induced radiation
  trapping},}\ }\href {\doibase 10.1103/PhysRevLett.104.023602} {\bibfield
  {journal} {\bibinfo  {journal} {Phys. Rev. Lett.}\ }\textbf {\bibinfo
  {volume} {104}},\ \bibinfo {pages} {023602} (\bibinfo {year}
  {2010})}\BibitemShut {NoStop}%
\bibitem [{\citenamefont {Cotrufo}\ and\ \citenamefont
  {Al{\`u}}(2018)}]{CotrufoAluarXiv18}%
  \BibitemOpen
  \bibfield  {author} {\bibinfo {author} {\bibfnamefont {M.}~\bibnamefont
  {Cotrufo}}\ and\ \bibinfo {author} {\bibfnamefont {A.}~\bibnamefont
  {Al{\`u}}},\ }\href@noop {} {\enquote {\bibinfo {title} {Single-photon
  embedded eigenstates in coupled cavity-atom systems},}\ } (\bibinfo {year}
  {2018}),\ \Eprint {http://arxiv.org/abs/1805.03287} {arXiv:1805.03287}
  \BibitemShut {NoStop}%
\bibitem [{\citenamefont {Wootters}(1998)}]{WoottersPRL98}%
  \BibitemOpen
  \bibfield  {author} {\bibinfo {author} {\bibfnamefont {W.~K.}\ \bibnamefont
  {Wootters}},\ }\bibfield  {title} {\enquote {\bibinfo {title} {Entanglement
  of formation of an arbitrary state of two qubits},}\ }\href {\doibase
  10.1103/PhysRevLett.80.2245} {\bibfield  {journal} {\bibinfo  {journal}
  {Phys. Rev. Lett.}\ }\textbf {\bibinfo {volume} {80}},\ \bibinfo {pages}
  {2245--2248} (\bibinfo {year} {1998})}\BibitemShut {NoStop}%
\bibitem [{\citenamefont {Shen}\ and\ \citenamefont {Fan}(2007)}]{ShenPRA07}%
  \BibitemOpen
  \bibfield  {author} {\bibinfo {author} {\bibfnamefont {J.-T.}\ \bibnamefont
  {Shen}}\ and\ \bibinfo {author} {\bibfnamefont {S.}~\bibnamefont {Fan}},\
  }\bibfield  {title} {\enquote {\bibinfo {title} {Strongly correlated
  multiparticle transport in one dimension through a quantum impurity},}\
  }\href {\doibase 10.1103/PhysRevA.76.062709} {\bibfield  {journal} {\bibinfo
  {journal} {Phys. Rev. A}\ }\textbf {\bibinfo {volume} {76}},\ \bibinfo
  {pages} {062709} (\bibinfo {year} {2007})}\BibitemShut {NoStop}%
\bibitem [{\citenamefont {Zheng}\ \emph {et~al.}(2010)\citenamefont {Zheng},
  \citenamefont {Gauthier},\ and\ \citenamefont {Baranger}}]{ZhengPRA10}%
  \BibitemOpen
  \bibfield  {author} {\bibinfo {author} {\bibfnamefont {H.}~\bibnamefont
  {Zheng}}, \bibinfo {author} {\bibfnamefont {D.~J.}\ \bibnamefont {Gauthier}},
  \ and\ \bibinfo {author} {\bibfnamefont {H.~U.}\ \bibnamefont {Baranger}},\
  }\bibfield  {title} {\enquote {\bibinfo {title} {{Waveguide {QED}: Many-body
  bound-state effects in coherent and {F}ock-state scattering from a two-level
  system}},}\ }\href {\doibase 10.1103/PhysRevA.82.063816} {\bibfield
  {journal} {\bibinfo  {journal} {Phys. Rev. A}\ }\textbf {\bibinfo {volume}
  {82}},\ \bibinfo {pages} {063816} (\bibinfo {year} {2010})}\BibitemShut
  {NoStop}%
\bibitem [{\citenamefont {Gonzalez-Ballestero}\ \emph
  {et~al.}(2013)\citenamefont {Gonzalez-Ballestero}, \citenamefont
  {Garcia-Vidal},\ and\ \citenamefont {Moreno}}]{GonzalezBallestroNJP13}%
  \BibitemOpen
  \bibfield  {author} {\bibinfo {author} {\bibfnamefont {C.}~\bibnamefont
  {Gonzalez-Ballestero}}, \bibinfo {author} {\bibfnamefont {F.~J.}\
  \bibnamefont {Garcia-Vidal}}, \ and\ \bibinfo {author} {\bibfnamefont
  {E.}~\bibnamefont {Moreno}},\ }\bibfield  {title} {\enquote {\bibinfo {title}
  {Non-{M}arkovian effects in waveguide-mediated entanglement},}\ }\href
  {\doibase 10.1088/1367-2630/15/7/073015} {\bibfield  {journal} {\bibinfo
  {journal} {New J. Phys.}\ }\textbf {\bibinfo {volume} {15}},\ \bibinfo
  {pages} {073015} (\bibinfo {year} {2013})}\BibitemShut {NoStop}%
\bibitem [{\citenamefont {Fang}\ and\ \citenamefont
  {Baranger}(2015)}]{FangPRA15}%
  \BibitemOpen
  \bibfield  {author} {\bibinfo {author} {\bibfnamefont {Y.-L.~L.}\
  \bibnamefont {Fang}}\ and\ \bibinfo {author} {\bibfnamefont {H.~U.}\
  \bibnamefont {Baranger}},\ }\bibfield  {title} {\enquote {\bibinfo {title}
  {{Waveguide QED: Power spectra and correlations of two photons scattered off
  multiple distant qubits and a mirror}},}\ }\href {\doibase
  10.1103/PhysRevA.91.053845} {\bibfield  {journal} {\bibinfo  {journal} {Phys.
  Rev. A}\ }\textbf {\bibinfo {volume} {91}},\ \bibinfo {pages} {053845}
  (\bibinfo {year} {2015})},\ \bibinfo {note}
  {\href{http://link.aps.org/doi/10.1103/PhysRevA.96.059904}{\textit{ibid.}
  \textbf{96}, 059904(E) (2017).}}\BibitemShut {Stop}%
\end{thebibliography}%
\nocite{FangCPC18,apsrev41Control}
\putbib[WQED,\jobname,REVTeX-longbib]
\end{bibunit}

\clearpage

% ----------------------------------------------------------------------------
% supplementary material starts
\begin{bibunit}[apsrev4-1]

\renewcommand{\bibnumfmt}[1]{[S#1]}
\renewcommand{\citenumfont}[1]{S#1}
\renewcommand{\theequation}{S\arabic{equation}}
\renewcommand{\thefigure}{S\arabic{figure}}
\renewcommand{\thepage}{S\arabic{page}}  
\renewcommand{\thesection}{S\arabic{section}}
\renewcommand{\thetable}{S\arabic{table}}
\setcounter{equation}{0}
\setcounter{figure}{0}
\setcounter{page}{1}
\setcounter{section}{0}
%\setcounter{subsection}{0}

%\begin{center}
%	\textbf{\large Supplementary Material for ``Exciting a Bound State in the Continuum through Multi-Photon Scattering plus Delayed Quantum Feedback''}\\
%	\vspace{15pt}
%	Giuseppe Calaj\'o, Yao-Lung L.\ Fang,  Harold U.\ Baranger, and Francesco Ciccarello\\
%	(Dated: January 13, 2019)\\
%\end{center}
%\begin{center}
%\begin{minipage}{5.0in}
%	\vspace*{0.1in}\tableofcontents
%\end{minipage}
%\end{center}
	\begin{CJK*}{UTF8}{} % instructed by  http://journals.aps.org/pra/authors/author-names-information
	\CJKfamily{bsmi}
	
	\title{Supplementary Material for ``Exciting a Bound State in the Continuum through Multiphoton Scattering Plus Delayed Quantum Feedback''}

	\author{Giuseppe Calaj\'o }
	%\thanks{Present address: ICFO-Institut de Ciencies Fotoniques, 08860 Castelldefels (Barcelona), Spain.}
	\affiliation{Vienna Center for Quantum Science and Technology, Atominstitut, TU Wien, Stadionallee 2, 1020 Vienna, Austria}
	\thanks{Present address: ICFO-Institut de Ciencies Fotoniques, 08860 Castelldefels (Barcelona), Spain. Contact email giuseppe.calajo@icfo.eu.}
	
	\author{Yao-Lung L. Fang (方耀龍)}
	\affiliation{Department of Physics, Duke University, P.O. Box 90305, Durham, North Carolina 27708-0305, USA}
	\affiliation{Computational Science Initiative, Brookhaven National Laboratory, Upton, NY 11973, USA}
	\thanks{Present address}
    	\affiliation{National Synchrotron Light Source II, Brookhaven National Laboratory, Upton, NY 11973, USA}
    	\thanks{Present address}
	
	\author{Harold U. Baranger}
	\affiliation{Department of Physics, Duke University, P.O. Box 90305, Durham, North Carolina 27708-0305, USA}
	
	\author{Francesco Ciccarello}
	\affiliation{Department of Physics, Duke University, P.O. Box 90305, Durham, North Carolina 27708-0305, USA}
\affiliation{Universit$\grave{a}$ degli Studi di Palermo, Dipartimento di Fisica e Chimica, via Archirafi 36, I-90123 Palermo, Italia}

	\affiliation{NEST, Istituto Nanoscienze-CNR, Piazza S. Silvestro 12, 56127 Pisa, Italia} 
	
	\date{January 13, 2019}
	
	%%%
	
	\begin{abstract}
		%In this Supplementary Material...
		\begin{minipage}{5.0in}
			%\begin{center}
			\vspace*{0.1in}\tableofcontents
			%\end{center}
		\end{minipage}
	\end{abstract}

	\maketitle
\end{CJK*} % for Chinese characters
\widetext

\section{Computational methods}

In order to solve for various wavefunctions in the waveguide-QED setups that we consider, we have taken two different routes: treating the 1D field as a continuum or discretizing it as an effective coupled-cavity array (CCA) and working in the middle of the band, where band-edge effects are minimized and the dispersion is close to linear. In the main text, results shown in \figs2 and 3(a)-(b) were obtained through the continuum approach, while the discrete approach was employed for producing \figs3(c), 4 and 5.

\subsection{Continuum approach}

For the continuum case, we follow the approach reported in \rref \cite{FangNJP18}: Starting from the Schr\"{o}dinger equation, by unfolding the half space and tracing out the two-photon part, we arrive at a (1+1)-dimensional \emph{delay} partial differential equation for the wavefunction $\psi(x, t)$ [joint amplitude for qubit and one photon at position $x$, c.f.\ \eq(5) of the main text], which we solve using a tailored finite-difference time-domain (FDTD) code that has multithreading support for reducing the computation time considerably \cite{FangCPC18}.\footnote{This FDTD code is open-sourced at \url{https://github.com/leofang/FDTD} and also available in \rref \cite{FangCPC18}.} 
The two-photon wavefunction can be constructed straightforwardly once $\psi$ is solved \cite{FangNJP18,FangCPC18}.
Our FDTD code provides the numerically exact solution to the dynamics in the two-excitation sector. 

For solving the two-photon scattering problem of interest, we send in a two-photon exponential wavepacket with the qubit initially unexcited (set \texttt{init\_cond=3} in the code). In the region $x<-a$, the exact solution to $\psi$ is given by
\begin{equation}
	\psi(x<-a, t) = A \left[\varphi_1(x-vt) e_0^{(2)}(t) + \varphi_2(x-vt) e_0^{(1)}(t) \right],
\end{equation}
where $e_0^{(i)}(t)$ is the qubit wavefunction in the one-excitation sector, solved by assuming the qubit is initially unexcited and the presence of an incident single-photon wavepacket $\varphi_i(x)$, and $A$ is the normalization constant for the two-photon wavepacket (see the main text). With this expression, the problem of computing for $\psi(x>-a, t)$ becomes well-defined and is solved by  FDTD \cite{FangCPC18}.

% Leo: because the notations were quite different between our earlier works and the present work, it would take too much space (and be too confusing) to re-introduce the set of notations used in the code and the associated analytical expressions. Therefore, in this version I only scratch the surface of the whole complexity.
%Harold: this is fine. 

% memory limitation of FDTD
Our FDTD code is mainly memory bound. For typical cases (such as Fig.~2) we need at least 128 GB of RAM. For the most extreme case that we explored [$\Gamma\tau\approx25$ in Fig.~3(a), corresponding to $m=k_0a/\pi=40$ with $\omega_0=10\Gamma$], we use a machine with 450+ GB of RAM in the local cluster. A rule of thumb for estimating the memory usage is to take $16(2N_x+n_x)N_y/1024^3$ (in GB), where $n_x$ is the number of FDTD steps for completing a round trip between the atom and the mirror; all input parameters are explained in detail in \cite{FangCPC18}. In the aforementioned most demanding case, both $N_x$ and $N_y$ are about $1.2\times10^5$ and $n_x\sim5000$ (such that one resonant wavelength has $n_x/m\sim100$+ FDTD steps, yielding an accuracy of order $10^{-4}$), so the $\psi$-array alone takes roughly 438 GB. 
Useful tricks to explore the large $\tau$ regime include reducing $\omega_0/\Gamma$ [since $\Gamma\tau=2m\pi/(\omega_0/\Gamma)$] and reducing $n_x/m$. For these reasons we set $\omega_0/\Gamma=20$ for the plots in Fig.~2 and Fig.~3(b) and $\omega_0/\Gamma=10$ for the plot in Fig.~3(a), where, as previously mentioned, we explored the regime of long delay time. In this way, with the same amount of memory, we were able to complete more FDTD steps at the expense of reduced accuracy (obviously careful checks were necessary).
Note that the larger $\tau$ the longer the cutoff time $t_f$ needs to be because of the multiple bounces of the photon field before escaping.  
%\Leo{(Harold, I remember you have a beautiful argument for why this is true. Do you want to elaborate it somewhere?)}

\subsection{Discrete approach}\label{DS}
An alternative approach to simulate  the two-photon scattering process 
%(semi)infinite waveguide  
consists in modeling it as a CCA comprised of a 1D arrangement of $N \gg1$ identical resonators of frequency $\omega_c$ with nearest-neighbour coupling rate $J$, with one resonator coupled to the qubit with coupling rate $g$. Hamiltonian (1) of the main text is thus discretized as
\begin{equation}\label{HCCA}
\hat H=\omega_c\sum_{n=1}^{N }\hat c_n^{\dagger}\hat c_n-J\sum_{n=1}^{N }(\hat c_n^{\dagger}\hat c_{n-1}+\hat c_{n-1}^{\dagger}\hat c_{n})+\omega_0\hat\sigma^{\dagger}\hat \sigma+
g(\hat c_{\frac{a}{\ell}}^{\dagger}\hat\sigma+\hat c_{\frac{a}{\ell}}\hat\sigma^{\dagger}),
\end{equation}
where $\hat c_n$ ($\hat c_n^\dag$) annihilates (creates) a photon on the $n$th cavity mode while $J=v/(2\ell)$ and $g=V/\sqrt{l}$ (here $l$ stands for the longitudinal size of each fictitious cavity). 
By introducing the momentum operators $\hat a_k=\frac{1}{\sqrt{N}} \sum_x   e^{i k x}  \hat a_x$, with $k\in \left]-\pi,\pi\right]$ (we  are implicitly rescaling the wavevector as $k:=kl$), the first two terms of \eq\eqref{HCCA}, which represent the free field Hamiltonian $\hat H_f$, can be arranged in the diagonal form $\hat H_{\rm f}= \sum_k \omega_k \hat c^\dag_k \hat c_k$, with the normal frequencies 
\begin{equation}\label{dis}
\omega_k=\omega_c-2J\cos k
\end{equation}  
forming a band of width $4J$  centered at the bare cavity frequency $\omega_c$. 
%In different limits, the Hamiltonian~\eqref{HCCA} captures the main physics of several  setups, ranging form linear waveguides to slow-light  waveguides~\cite{CalajoPRA16}. In our case we are interested  in recovering the linear continuum waveguide limit. 
The dispersion law \eqref{dis} is approximately linear for $\omega\simeq \omega_c$. Thus, in the weak-coupling regime $\Gamma\ll 4J$ and for $\omega_c=\omega_0$, Hamiltonian~\eqref{HCCA} is a reasonable approximation to the physics of the continuous-waveguide Hamiltonian~(1) of the main text. The advantage of using this discrete model is that it can  be easily handled numerically up to the third-excitation sector of the Hilbert space.  
In particular, in this work we simulate the time-dependent Schr\"odinger equation corresponding to \eq\eqref{HCCA} for
%\begin{equation}
%|\Phi(t)\rangle=\sum_{n=1}^Nb_n(t)\hat c_n^{\dagger}|e,0\rangle+\frac{1}{\sqrt{2}}\sum_{n,m=1}^Nu_{n,m}(t)\hat c_n^{\dagger}\hat c_m^{\dagger}|g,0\rangle
%\end{equation}
$N \sim 800$ ($N\sim 100$) cavities for the two-(three-) excitation subspace and set $\Gamma/(4J)=(g^2/J)/(4J)\le 0.1$ in order to be consistent with the weak-coupling assumption. In particular we have used $\Gamma/(4J)=0.075$ for Figs. 3(c) and 5 and $\Gamma/(4J)=0.062$ for Fig. 4.

It is worth mentioning that, in the present finite-band CCA, even bound states \emph{outside} the band can occur~\cite{CalajoPRA16} (in \rref~\cite{LongoPRL10} it was shown that they can be populated via two-photon scattering).  However, the assumption $\omega_0=\omega_c$ ($k_0=\pi/2$) ensures that the BIC lies exactly at the center of the band in a way that processes that excite the bound states outside the continuum are energetically forbidden (see~\cite{CalajoPRA16}). 

Finally, note that the present discrete approach is straightforwardly extended to describe the two-qubit setup in \fig5 of the main text. In this case, we place the two emitters far from the CCA edges to avoid unwanted back-reflection from the array's ends within the considered simulation time.

\section{Derivation of \eq (8)}

The BIC state has the structure (see main text) 
\begin{align}
|\phi_b\rangle=\varepsilon_b |e,0\rangle+\int_0^a\!{\rm d}x\,\left(f_R(x)\hat a_R^\dag(x)+f_L(x)\hat a_L^\dag(x)\right)|g,0\rangle\label{phib_sm}
\end{align}
where
\begin{align}
|\varepsilon_b|^2+\int_0^a\!{\rm d}x\,|f_R(x)|^2+\int_0^a\!{\rm d}x\,|f_L(x)|^2=1, \qquad |\varepsilon_b|^2=\frac{1}{1+\tfrac{1}{2} \Gamma\tau}\,.\label{norm}
\end{align}
Based on definitions (5) and (6) of the main text, we thus find that in the case of state (7)
\begin{equation}
P_{\rm e}(\infty)=\int_a^\infty \!{\rm d}x\,|\xi(x,t)|^2|\varepsilon_b|^2\,, \qquad P_{\rm ph}(\infty)=\int_a^\infty \!{\rm d}x|\xi(x,t)|^2 \sum_{\eta=R,L}\int_0^a \!{\rm d}y\,|f_\eta(y)|^2\label{Pe} \,,
\end{equation}
where it is understood that time $t$ is large enough that the scattering is complete.
Hence, due to the normalization condition in \eqs\eqref{norm},
\begin{equation}
P_{\rm tr}(\infty)=P_{\rm e}(\infty)+P_{\rm ph}(\infty)=\int_a^\infty \!{\rm d}x\,|\xi(x,t)|^2=P_{\rm BIC}\,\,.
\label{eq: Ptr PBIC}
\end{equation}
Next, combining the last identity in \eqref{norm} with the first of \eqs\eqref{Pe} yields $\int_a^\infty \!{\rm d}x\,|\xi(x,t)|^2={P_{\rm e}}(\infty)/{|\varepsilon_b|^2}=(1{+}\tfrac{1}{2} \Gamma\tau)P_{\rm e}(\infty)$. Thereby,
\begin{equation}
P_{\rm tr}(\infty)=P_{\rm BIC}=(1 + \tfrac{1}{2} \Gamma\tau)P_{\rm e}(\infty)\,,
\label{eq: Ptr PBIC Pe}
\end{equation}
which completes the proof of \eq(8) in the main text.

\section{Optimal width for resonant two-photon wavepackets}

\begin{figure}[h]
	\centering
	\includegraphics[width=0.45\textwidth]{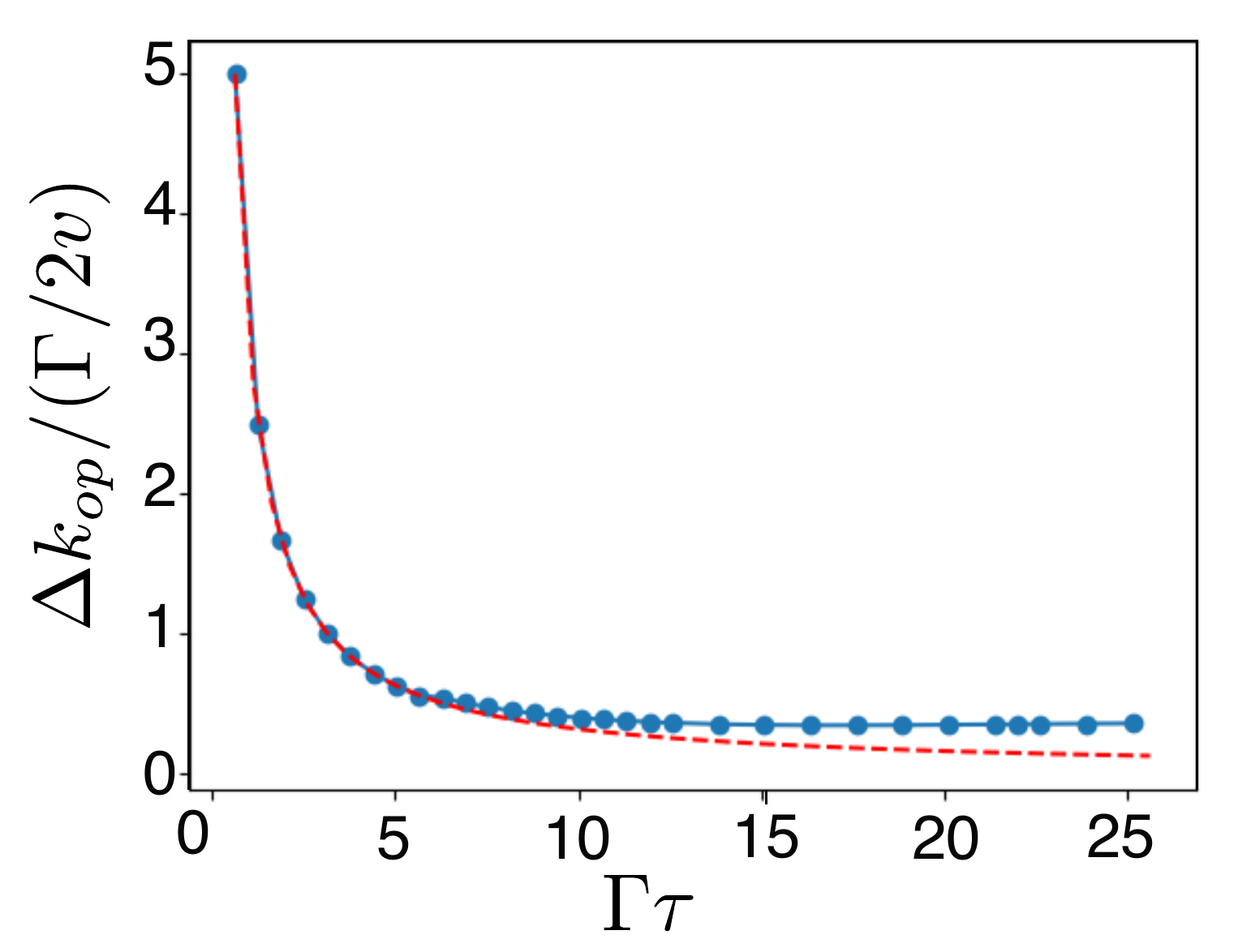}
	\caption{Optimal wavepacket width $\Delta k_{op}$ as a function of $\Gamma\tau$, which was used to produce \fig3(a) in the main text. 
		%For small $\Gamma\tau$ we find that an empirical formula $\Delta k_{op}=\pi/(2v\tau)$ (red dashed line) fits the result quite well. However, $\Delta k_{op}$ saturates for large $\Gamma\tau$, in order that the qubit continue to be excited by the first passage of the injected pulse.  
		The behavior in the range $\Gamma\tau\lesssim2\pi$ is well fitted by the function $\Delta k_{op}=\pi/(2v\tau)$ (red dashed line).  Here we set $\omega_0=10\Gamma$.
		\label{fig: optimal alpha}}
\end{figure}

In \fig3(a) of the main text, we plot various probabilities against $\Gamma\tau$, each obtained after an optimization over the wavepacket width $\Delta k$. In Fig.~\ref{fig: optimal alpha}, we plot the optimal $\Delta k$, denoted by $\Delta k_{op}$, as a function of $\Gamma\tau$. We find that, for delays such that $\Gamma\tau\lesssim2\pi$, $\Delta k_{op}$ scales approximately as $\Delta k_{op}\approx\pi/(2v\tau)=\pi/(4a)$. For larger delays, $\Delta k_{op}$ instead saturates to a non-zero value (presumably in order to maximize the atomic absorption during the scattering transient).
% \Leo{(do we have better argument? I am taking the similar explanation used in the main text...)}. 
% Harold: this seems fine to me. 
For each set value of $\Gamma\tau$, the optimization was carried out by interpolating the $P_{\rm e}$'s data computed on a discrete $\Delta k$-grid (on a log scale) and working out the $\Delta k$ at which the curve $P_{\rm e}(\Delta k)$ exhibits a local maximum. 

\section{Increasing the BIC generation probability}

\fig4 of the main text reports a paradigmatic example showing that $P_{\rm tr}(\infty)$ can be strongly increased (compared to the standard choice of an exponential pulse) by engineering the wavepacket shape. Here, we describe how the wavepacket in \fig4 was derived.

The BIC generation process would be perfect ($P_{\rm BIC}=1$) if, in \eq(7) of the main text, $\beta_{RR}(x,y,t_f)=0$, meaning that the {\it incoming} two-photon wavepacket deterministically evolves at the end of scattering into the BIC plus a (normalized) {outgoing} single-photon wavepacket $\xi_R(x,t)$.
The basic idea is to consider the time-reversed version of such ideal process: the system is initially prepared in the dressed BIC $|\phi_b\rangle$ and an {incoming} single-photon wavepacket $\xi_{\rm op}(x,t)$ undergoes scattering resulting in a normalized {\it outgoing} two-photon wavepacket $\beta_{RR,{\rm op}}(x,y,t)$ at the end of scattering \cite{CotrufoAluarXiv18}. If such a single-photon wavepacket $\xi_{{\rm op}}(x,t)$ exists, then sending on the initially unexcited qubit the time-reversed version of $\beta_{RR,{\rm op}}(x,y,t)$ will deterministically yield the BIC (plus an outgoing photon). 

Based on the above, we considered a right-incoming, Gaussian, single-photon wavepacket of the form
\begin{figure}[h]
	\centering
	\includegraphics[width=0.75\textwidth]{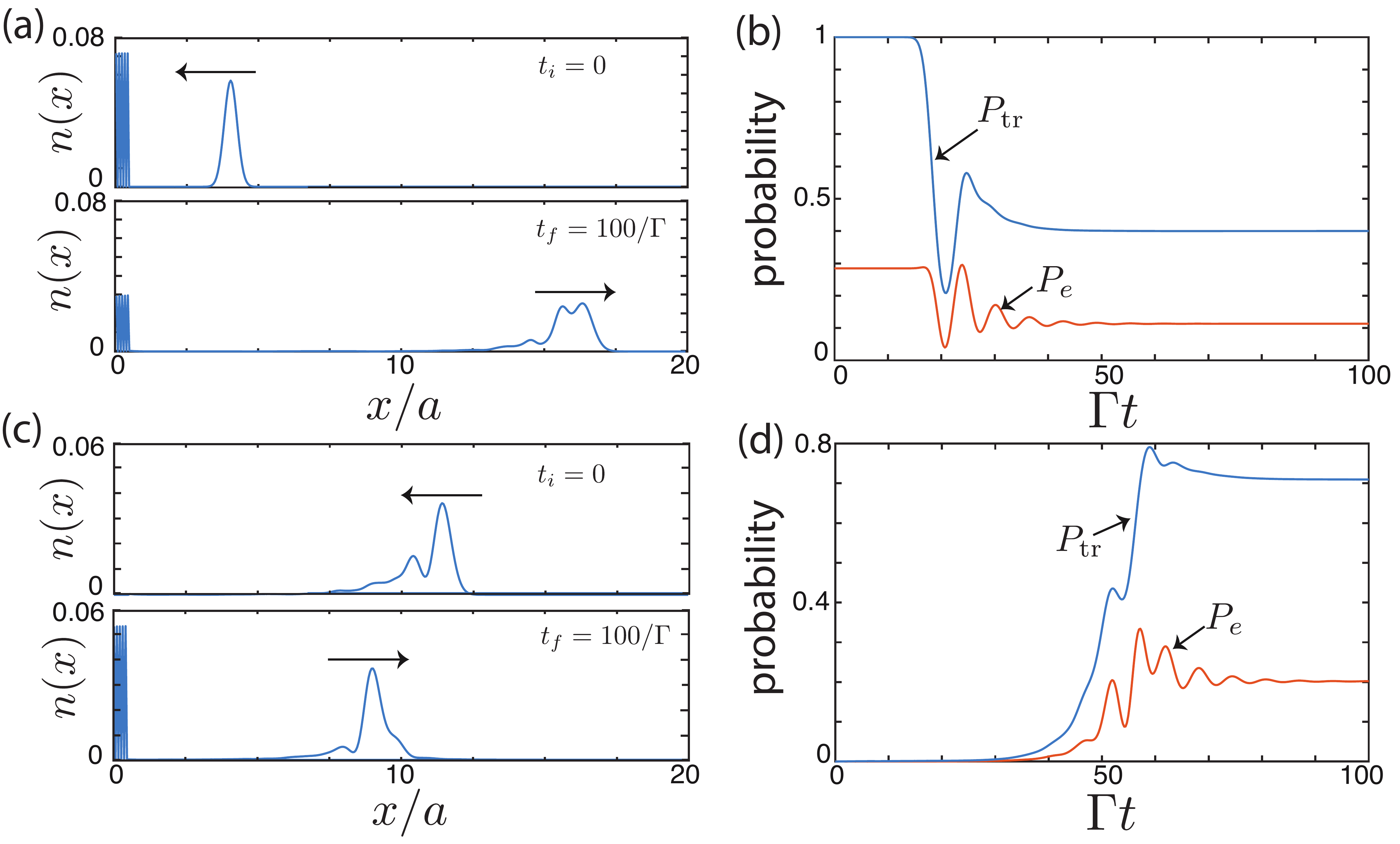}
	\caption{ (a)-(b): Scattering of a single-photon wavepacket with wavefunction \eqref{gaussian} for $\Delta x=2v/\Gamma$ when the system is prepared in the BIC $|\phi_b\rangle$. (a): Photon-density profile before (top panel) and after (bottom panel) scattering. (b): Time behavior of $P_{\rm tr}$ and $P_{\rm e}$ (see main text) in the same process. (c)-(d): Scattering of the two-photon wavepacket obtained by time-reversing and normalizing the two-photon outgoing component of the previous process. (c):  Photon-density profile before and after scattering. (d): Time behavior of $P_{\rm tr}$ and $P_{\rm e}$ in the same process. Throughout, we set  $k_0a{=}20\pi$ and $\Gamma\tau=5$. 
	\label{fig_time_rev}}
\end{figure}
\begin{equation}\label{gaussian}
\xi_{L}(x)=\frac{1}{\sqrt[4]{\pi\Delta x^2}}\,e^{-\frac{(x-x_0)^2}{2\Delta x^2}-ik_0(x-x_0)},
\end{equation}
where $\Delta x$ is the spatial width and $|x_0-a|>3\Delta x$, which is injected toward  the qubit-mirror region initially prepared in the BIC [see \fig\ref{fig_time_rev}(a)]. At the end of scattering, as shown in Fig.~\ref{fig_time_rev}(b), part of the excitation trapped in the BIC is released, while a residual amount remains. Minimizing the latter over the width $\Delta x$ yields $\Delta x =2v/\Gamma$. The final joint state is of the same form as \eq(7) in the main text, featuring in particular an outgoing two-photon component. We normalize and time-reverse this two-photon wavefunction, which is then used as the input of our original problem (namely, it is sent to the scattering region with the qubit initially in the ground state) as shown in Fig.~\ref{fig_time_rev}(c). It turns out [see Fig.~\ref{fig_time_rev}(d)] that the resulting two-photon-wavepacket is far more effective than the exponential one (see \figs 2 and 3 in the main text), leading to the BIC generation probability $P_{\rm BIC}=P_{\rm tr}(\infty)\simeq 70\%$. 

A further improvement can be obtained by repeating the above process, but this time choosing $\xi_\text{op}(x,t)$ as the single-photon outgoing component of Fig.~\ref{fig_time_rev}(c) (once this is normalized and time-reversed). One more iteration of this procedure yields the results in \fig5 of the main text, where we obtain in particular $P_{\rm tr}(\infty)\simeq 80\%$ (further iterations do not substantially change this value).
%This instance shows that, in principle, with a properly shaped wave packet it should be possible to excite the BIC trough non-linear scattering with probability close to one. 

In this paradigmatic instance, we started from a Gaussian-shaped single-photon wavepacket. Deriving a systematic, general optimization criterion that does not rely on such a specific choice is the target ongoing investigation, in particular with the goal to assess whether the BIC generation probability can approach 100\%.

\section{Two-qubit BIC}

In \fig5(b) of the main text we plot $P_{\rm tr}$, $P_{\rm e}$ and the concurrence $C$ for the two-qubit setup of \fig5(a). Here, we provide detailed definitions of these quantities.

As in the case of a qubit in front of a mirror [\cf\eq(5) in the main text], the joint state at time $t$ lives in the two-excitation sector and thus has the general form 
\begin{equation}\label{state_BIC_2atom}
\begin{split}
	|\Psi(t)\rangle=&
\left( f(t)\hat \sigma_1^\dag\hat \sigma_2^\dag +\sum_{\eta=R,L}\int_{-\infty}^{\infty}\!\!{\rm d}x\, \psi_{1\eta}(x,t) \hat a_\eta^{\dagger}(x)\hat \sigma_1^\dag +\sum_{\eta=R,L}\int_{-\infty}^{\infty}\!\!{\rm d}x\, \psi_{2\eta}(x,t) \hat a_\eta^{\dagger}(x)\hat \sigma_2 ^\dag
\right.\\
 &  \left.   	+\sum_{\eta,\eta'=R,L}\tfrac{1}{\sqrt{2}}\iint_{-\infty}^{\infty}\!\! {\rm d}x {\rm d}y\,
\chi_{\eta,\eta'}(x,y,t) \,\hat a_\eta^{\dagger}(x) \hat a_{\eta'}^{\dagger}(y)) \right)   |g_1,g_2\rangle|0\rangle\,.
	\end{split}
	\end{equation}
	Accordingly, the total probability that at least one qubit is excited reads
%in manner similar to the atom-mirror case:
	\begin{equation}
	P_{\rm e}(t) = |f(t)|^2+\sum_{\eta}\!\int_{-\infty}^{\infty}\!\!{\rm d}x \,|\psi_{1\eta}(x,t)|^2+\sum_{\eta}\!\int_{-\infty}^{\infty}\!\!{\rm d}x \,|\psi_{2\eta}(x,t)|^2,\label{Pe2at}
	\end{equation}
	while the probability that one photon lies between the two qubits and one beyond them is given by
	\begin{equation}
	P_{\rm ph}(t) = 2\sum_{\eta,\eta'}\int_{-a}^{a}{\rm d}x \int_{-\infty}^{-a}{\rm d}y \,
|\chi_{\eta,\eta'}(x,y,t)|^2+2\sum_{\eta,\eta'}\int_{-a}^{a}{\rm d}x \int_{a}^{\infty}{\rm d}y \,
|\chi_{\eta,\eta'}(x,y,t)|^2\,,
	\end{equation}
	while their sum is called $P_{\rm tr}=P_{\rm e}+P_{\rm ph}$.
	
	To measure the amount of qubit-qubit entanglement in state \eqref{state_BIC_2atom}, we trace out the field degrees of freedom to get the reduced two-qubit density matrix and next calculate the corresponding Wootters concurrence \cite{WoottersPRL98}. This takes the form
	\begin{equation}
	C(t)= \max \left(0, 2|C_{12}(t)|-\sqrt{|f(t)|^2 P_\textrm{ph}}\right),
	\end{equation}
	where
	\begin{equation}
	C_{12}(t){=}\sum_{\eta}\!\int_{-\infty}^{\infty}\!\!{\rm d}x \,\psi_{1\eta}^*(x,t)\psi_{2\eta}(x,t) \label{coh2at}
	\end{equation}
	are the atomic coherences.

\section{Resilience to qubit losses into an external reservoir}
In the main text, we considered ideal setups in which each qubit is perfectly coupled to the waveguide. In practice, the emitters may be also in contact with modes other than the waveguide, which introduces additional decay channels. Assuming a Markovian external reservoir and given our purpose of describing a scattering process, such dissipation can be accounted for by adding to the Hamiltonian \eq(1) of the main text an effective non-Hermitian term \cite{ShenPRA07,ZhengPRA10} (we consider the setup of \fig1 in the main text):
\begin{equation}\label{Hdiss}
\hat H\rightarrow \hat H-i\frac{\gamma_a}{2}\,\hat\sigma^{\dagger}\hat\sigma
\end{equation}
where $\gamma_a$ is the decay rate of the qubit  into the external environment.
\begin{figure}[h]
	\centering
	\includegraphics[width=0.50\textwidth]{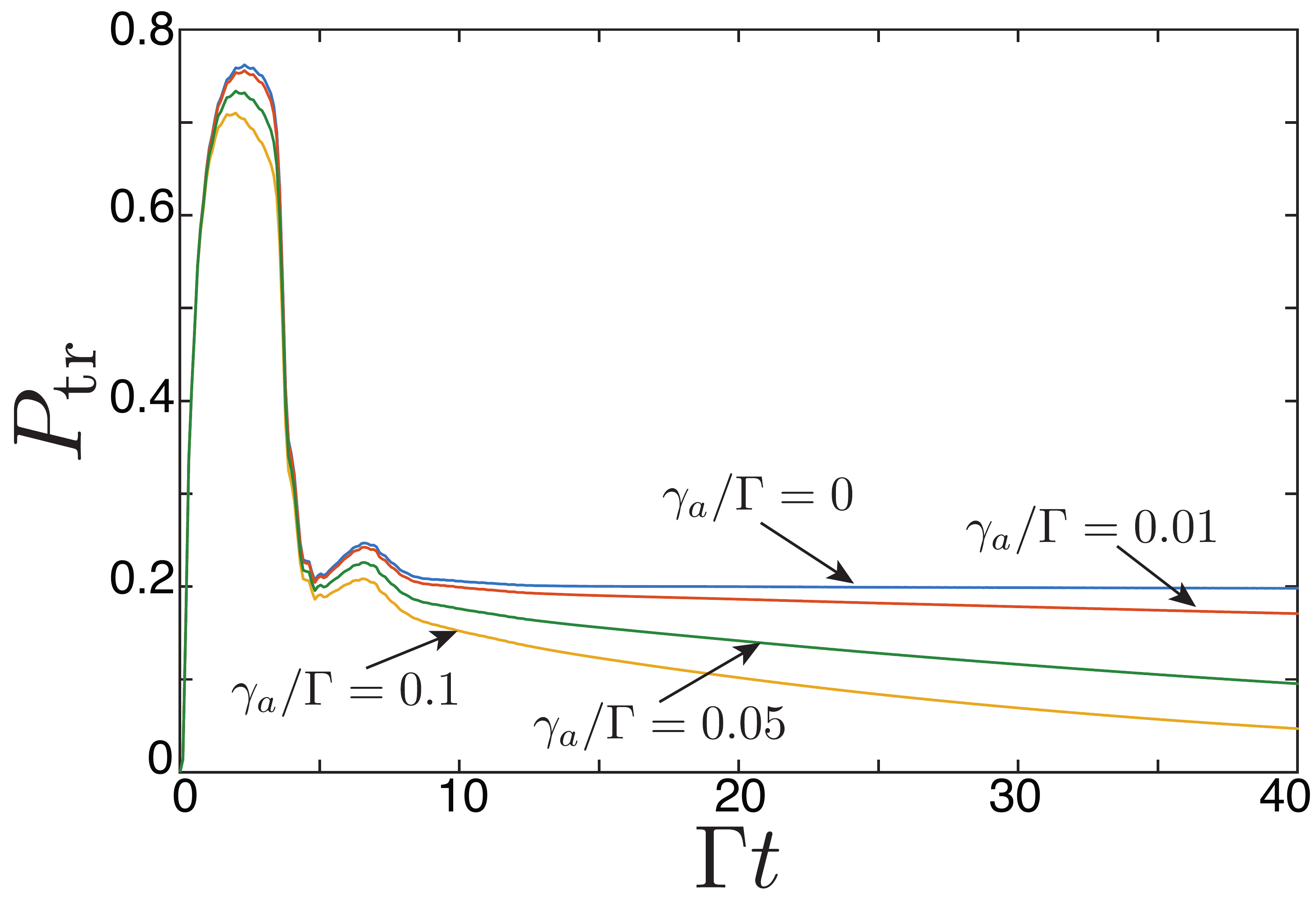}
	\caption{Trapping probability $P_{\rm tr}$ is plotted against time for different loss rates, $\gamma_a$, of the qubit into an external environment  for the one-qubit setup of \fig1 in the main text. The computation was done using the discrete approach as outlined in Sec.~\ref{DS} with $\Gamma/(4J)=0.04$. Other parameters are the same as in Fig.~2.\label{fig: decay}}
	%and considering a two photon exponential packet (as in the main text) with parameters $\Gamma\tau=\pi$, $k_0a=10\pi$,   and $\Delta k=\Gamma/(2v)$.}
\end{figure}

In \fig\ref{fig: decay}, we use the same parameters as in \fig2 of the main text and study how the time dependence of $P_{\rm tr}$ is affected by the loss rate $\gamma_a$ (recall that for $\gamma_a=0$ the asymptotic value of $P_{\rm tr}$ is the probability of generating the BIC, $P_{\rm BIC}$). In the range considered ($\gamma_a$ up to 10\% of $\Gamma$, the decay rate into the waveguide modes) the presence of a nonzero loss indeed turns the long-time saturation at $P_{\rm BIC}$ into a slow decay. We found strong numerical evidence that the long-time behavior of $P_{\rm tr}$ is described by the function $P_{\rm tr}(t\gg\tau)\sim P_{\rm tr}(\infty)e^{-\gamma_a |\varepsilon_b|^2t}$ with $P_{\rm tr}(\infty)=P_{\rm BIC}$ for $\gamma_a=0$ [see \eqs(8), \eqref{eq: Ptr PBIC} and \eqref{eq: Ptr PBIC Pe}] and $|\varepsilon_b|^2$ being the emitter's population in the BIC [see \eqs(3) and (4) in the main text].
This finding can be heuristically explained through the following simple argument. %Including the extra decay corresponds to adding to the Hamiltonian
The non-Hermitian term in \eq\eqref{Hdiss} can be equivalently expressed as $-i \frac{\gamma_a}{2} |e\,0\rangle\langle e\,0|$, where we set $|e\,0\rangle=|e\rangle|0\rangle$. The effective representation of this in the (one-dimensional) BIC subspace is obtained upon projection as 
%$$|\phi_b\rangle\langle \phi_b| (\hat H-i \gamma_a |e\,0\rangle\langle e\,0|)|\phi_b\rangle\langle \phi_b|=\omega_0 |\phi_b\rangle\langle \phi_b|-i\gamma_a |\langle e\,0|\phi_b\rangle|^2$$
\begin{equation}
|\phi_b\rangle\langle \phi_b| \left(-i \frac{\gamma_a}{2}  |e\,0\rangle\langle e\,0|\right)|\phi_b\rangle\langle \phi_b|=-i\frac{\gamma_a}{2}  |\langle e\,0|\phi_b\rangle|^2\,|\phi_b\rangle\langle \phi_b|=-i\frac{\gamma_a}{2} |\varepsilon_b|^2\,|\phi_b\rangle\langle \phi_b|\,,\,
\end{equation}
indicating that for $\gamma_a\neq 0$ the BIC decays at rate $\gamma_a |\varepsilon_b|^2$.

It follows that the smaller the excitonic fraction $|\varepsilon_b|^2$, the more robust is the BIC to emitter losses. Since $|\varepsilon_b|^2$ decreases with $\tau$ [\cf\eq(3) in the main text] it turns out that the longer the delay, the more resilient is the BIC generation scheme to loss. This thus embodies a further significant advantage compared to generation schemes based on spontaneous emission (see discussion in the main text), especially if one aims at an almost pure single-photon trapping.

A detailed discussion of the effect of qubit losses on single-photon BICs, which explicitly takes into account the degrees of freedom of the extra reservoir, is presented in \rref\cite{GonzalezBallestroNJP13} (in particular it is shown that the external bath dresses the BIC and scattering states without affecting their orthogonality).

\section{Role of emitter nonlinearity in the BIC generation scheme}

In the main text, we stressed that the qubit nonlinearity is key to generate the BIC. In this section, we further illustrate this point in terms of a bosonic Hamiltonian that depends in particular on the strength of nonlinearity \cite{LongoPRL10,FangPRA15}. For the sake of argument, we focus again on the one-qubit setup of \fig1. 

Consider the bosonic Hamiltonian 
\begin{align}\label{Hbos}
	\hat {\cal H}(U)=\hat H_{\sigma\rightarrow b}+ \hat H_I(U)\,\,\,\,\,{\rm with}\,\,\,\,\hat H_I(U)= \frac{U}{2}\hat b^\dagger\hat b\,(\hat b^\dagger\hat b-1)\,\,.
	\end{align}
Here, $\hat H_{\sigma\rightarrow b}$ is in the same form as $\hat H$ in \eq(1) of the main text except for the replacement $\hat{\sigma}\rightarrow \hat{b}$, where $\hat{b}$ and $\hat{b}^\dag$ fulfill bosonic commutation relations $[\hat{b}, \hat{b}^\dag]=1$ and commute with all the waveguide operators. The {\it quartic} term $\hat H_I(U)$ describes a fictitious, on-site boson-boson repulsion at the emitter's location. When this is absent, \ie for $U=0$, \eq\eqref{Hbos} is equivalent to replacing the qubit with a bosonic mode, which makes the total Hamiltonian fully bosonic and quadratic. For $U\neq0$, the effect of $\hat H_I(U)$ is to make energetically unfavorable the occupation of emitter's number states $|n\rangle=(\hat b^\dag)^n|{\rm vac}\rangle/\sqrt{n!}$ with $n\ge 2$ so that in the limit of infinite $U$ this behaves as an effective qubit and we recover the Hamiltonian  \eq(1):
\begin{equation}
\hat H=\lim_{U\to \infty}\hat {\cal H}(U)\label{H02}\,.
\end{equation}
We next discuss the cases $U=0$ and $U\neq 0$ separately.

\subsection{$U=0$}

In this case, the Hamiltonian is bosonic and quadratic and can thereby be diagonalized straightforwardly in terms of the normal modes. In the single-excitation sector $N=1$, where $\hat N=\hat b^\dag\hat b+\sum_{\eta=R,L}\int\!{\rm d}x\hat a_{\eta}^\dag(x)\hat a_{\eta}(x)$, the eigenstates of \eqref{Hbos} consist of scattering states $\{|\phi_k\rangle\}$ and a bound state $|\phi_b\rangle$ such that $\hat H_{\sigma\rightarrow b}|\phi_k\rangle=\omega_k  |\phi_k\rangle$ and $\hat H_{\sigma\rightarrow b}|\phi_b\rangle=\omega_0  |\phi_b\rangle$, respectively. The corresponding wavefunctions are given by $|\phi_j\rangle=\hat \alpha_j^\dag |{\rm vac}\rangle$, where $|{\rm vac}\rangle=|g\rangle |0\rangle$ and the normal-mode operators $\{\hat\alpha_j\}$ are defined as
\begin{eqnarray}
\hat \alpha_j=\varepsilon_j^* \,b+\!\sum_{\eta=R,L}\int_0^\infty\!{\rm d}x \,f^*_{j,\eta}(x)\,a_{\eta}(x)%\,\,\,\,\,\,\,\,\,\,(j=k,b)\,.
\end{eqnarray}
with $j=k, b$.
Here, the amplitudes $\{\varepsilon_j\}$ and $\{f_{j,\eta}(x)\}$ are the same as those defining the corresponding single-excitation stationary states of the one-qubit Hamiltonian $\hat H$. Regardless of $N$, normal-mode operators $\{\hat\alpha_j\}$ allow one to express the Hamiltonian in the diagonal form
\begin{equation}
\hat {\cal H}(U=0)
=\hat{H}_{\sigma\rightarrow b}
=\int_0^\infty\!{\rm d}k \,\omega_k\, \hat \alpha_k^\dag \hat \alpha_k+\omega_0 \hat \alpha_b^\dag \hat \alpha_b.
\end{equation}

In the two-excitation sector, corresponding to $N=2$, the eigenstates read
\begin{eqnarray}
|1_k1_{k'}\rangle=\frac{1}{\sqrt 2}\hat \alpha_k^\dag\hat \alpha_{k'}^\dag |{\rm vac}\rangle\,,\,\,\,\,|1_k 1_{b}\rangle=\hat \alpha_k^\dag\hat \alpha_{b}^\dag |{\rm vac}\rangle\,,\,\,\,\,|2_b\rangle=\frac{1}{\sqrt 2}\left(\hat \alpha_b^\dag\right)^2 \!|{\rm vac}\rangle\,,\label{HSS2}
\end{eqnarray}
with eigenvalues $\omega_k+\omega_{k'}$, $\omega_k+\omega_{0}$ and $2\omega_0$, respectively. These are physically interpreted as follows:
\begin{itemize}
	\item[-] states $|1_k1_{k'}\rangle$ are two-photon scattering states describing two incoming photons that scatter off the emitter and mirror and are eventually fully reflected;
	\item[-] states $|1_k1_{b}\rangle$ are semi-bound states: one photon scatters off while another photon is confined between the emitter and mirror dressing the emitter so as to form the single-excitation bound state $|\phi_b\rangle$;
	\item[-] state $|2_{b}\rangle$ is a two-photon bound state, featuring in particular two photons fully confined within the mirror-emitter interspace.
\end{itemize}
Based on these, we see that the scheme for generating the BIC via two-photon scattering becomes fully ineffective when the qubit is replaced by a bosonic mode: injecting two photons with the emitter initially unexcited involves only the stationary states $|1_k1_{k'}\rangle$, which have no overlap with other eigenstates $\{|1_k1_b\rangle\}$ and $|2_b\rangle$. The emitter will typically be excited during the scattering transient, but the two photons will be eventually fully reflected with no light confined in the mirror-emitter interspace. This conclusion holds regardless of the parameters, hence in particular no matter how long the time delay.

\subsection{$U\ne 0$}

The last conclusion does not hold any more when the fictitious bosonic repulsive term $\hat H_I(U)$ is present [\cf\eq\eqref{Hbos}].
In the two-excitation sector $N=2$, this takes the effective form $\hat H_I^{(N=2)}(U)=U\,|bb\rangle\langle bb|$ with $|bb\rangle=(\hat b^\dag)^2|{\rm vac}\rangle/\sqrt{2!}$. Noting now that in the same subspace all the stationary states [see \eq(\ref{HSS2})] generally feature a term $\propto|bb\rangle$ we see that the repulsive interaction $\hat H_I(U)$ mixes together all the eigenstates of $\hat H_{\sigma\rightarrow b}$. Most importantly, this means that introducing the nonlinearity has in particular the effects of (i) {\it connecting} two-photon scattering states $\left\{|1_k 1_{k'}\rangle\right\}$ to semi-bound ones $\left\{|1_k 1_{b}\rangle\right\}$ [see \eq(\ref{HSS2})] and (ii) \emph{eliminating}, as $U\rightarrow\infty$, the two-photon bound state $|2_b\rangle$. 

Indeed, using the Lippmann-Schwinger formalism, it can be shown that the $N=2$ unbound stationary states of $\hat{\cal H}(U)$ have the following form \cite{FangPRA15}
\begin{equation}
	|\psi_2(k_1, k_2)\rangle = |1_{k_1}1_{k_2}\rangle +\frac{U\langle bb |1_{k_1}1_{k_2}\rangle }{1-UG_{bb}(E)} \hat{G}^R(E)|bb\rangle,
	\label{eq: finite U N=2 eigenstate}
\end{equation}
where $\hat{\cal H}(U)|\psi_2(k_1, k_2)\rangle=E|\psi_2(k_1, k_2)\rangle$ with $E=\omega_{k_1}+\omega_{k_2}$, $\hat{G}^R(E)=(E-\hat{H}_{\sigma\rightarrow b}+i\delta)^{-1}$ is the retarded Green's function for $U=0$, and $G_{bb}(E)=\langle bb|\hat{G}^R(E)|bb\rangle$. Overlaps between the eigenstate $|\psi_2\rangle$ and other states can be easily computed by noting that all matrix elements can be expressed in terms of the amplitudes $\{\varepsilon_j\}$ and $\{f_{j,\eta}(x)\}$ from the single-excitation sector \cite{FangPRA15}. In the limit of $U\rightarrow\infty$, Eq.~\eqref{eq: finite U N=2 eigenstate} becomes
\begin{equation}
|\psi_2(k_1, k_2)\rangle = |1_{k_1}1_{k_2}\rangle - \frac{\langle bb |1_{k_1}1_{k_2}\rangle}{G_{bb}(E)} \hat{G}^R(E)|bb\rangle, 
\label{eq: infinite U N=2 eigenstate}
\end{equation}
which in particular entails $\langle bb | \psi_2\rangle = 0$. Thus, it is clearly impossible to doubly occupy the emitter, and the 2LS behavior is correctly recovered.

This argument provides further physical intuition as to why the nonlinearity of the 2LS, in this picture encoded in the repulsive term $\hat H_I(U)$, is essential for the BIC generation scheme via two-photon scattering. In particular, it shows how the intrinsic qubit nonlinearity enables the process where two incoming photons can evolve with some probability into a single scattering photon and the single-photon bound state $|\phi_b\rangle=|1_b\rangle$.

\begin{figure}[t]
	\centering
	\includegraphics[width=0.50\textwidth]{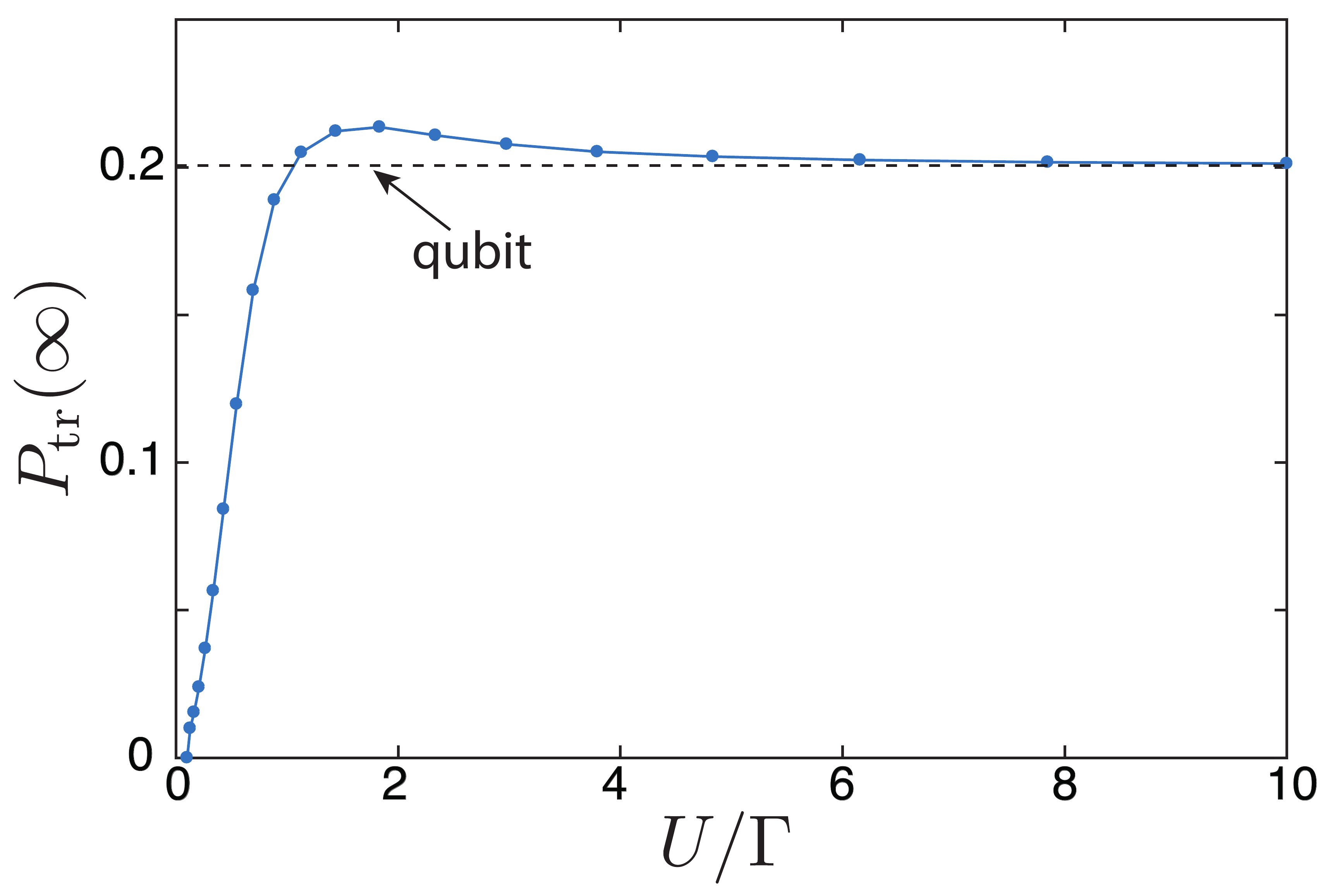}
	\caption{Asymptotic trapping probability, $P_{\rm tr}(\infty)$, as a function of the nonlinearity parameter $U$ in the case of two photons scattering off a {\it bosonic} emitter and a mirror (setup analogous to \fig1 in the main text) assuming the total Hamiltonian~\eqref{Hbos}. The black dashed line marks the value of $P_{\rm tr}(\infty)$ when the emitter is a qubit and the total Hamiltonian is the one in \eq(1) of the main text. The numerical method and parameters are the same as in \fig\ref{fig: decay}.}\label{fig_nl}
\end{figure}

To further support the above conclusion, using the method of Sec.~\ref{DS} we numerically computed in a paradigmatic case the asymptotic trapping probability $P_{\rm tr}(\infty)$ for the bosonic Hamiltonian~\eqref{Hbos} as a function of the nonlinearity parameter $U$.  The definition of $P_{\rm tr}$ is formally analogous to the one where the emitter is a qubit except that the probability of finding the emitter doubly excited (population of state $|bb\rangle$) is now also included. As shown in \fig\ref{fig_nl}, $P_{\rm tr}(\infty)$ {\it vanishes} when $U=0$ (bosonic emitter) and then overall grows with $U$, eventually converging as $U\rightarrow \infty$ to the corresponding value obtained with a qubit. 

As a final remark, we mention that, while for $U=0$ multi-photon BICs do exist and are simply given by $(\hat \alpha_b^\dag)^n/\sqrt{n!}\,|{\rm vac}\rangle$, for the two-level emitter addressed in the main text (corresponding to $U\rightarrow \infty$) in the range of parameters considered in this work we did not find any numerical evidence of their existence, which can be checked simply by examining the conservation of probability.

\nocite{apsrev41Control}
\putbib[WQED,REVTeX-longbib]
\end{bibunit}

\end{document}